\documentclass{aa}

\usepackage{newtxtext,newtxmath}
\usepackage[T1]{fontenc}
\usepackage[dvipsnames]{xcolor}
\usepackage{wasysym}
\usepackage{graphicx}
\usepackage{hyperref}  
\usepackage{xspace}
\usepackage{caption}
\usepackage{placeins}
\usepackage{xcolor}
\usepackage{subcaption}

\definecolor{afcolor}{HTML}{b3443c}

\begin{document}
\title{Physically motivated modeling of LyC escape fraction during reionization}

\author{Ivan Kostyuk\inst{1, 2}, Benedetta Ciardi\inst{1}, Andrea Ferrara\inst{2}}

\institute{Max-Planck-Institut f\"{u}r Astrophysik, Karl-Schwarzschild-Str. 1, 85741 Garching, Germany \\ {E-mail: ivkos@mpa-garching.mpg.de} \and Scuola Normale Superiore, Piazza dei Cavalieri 7, I- 50126 Pisa, Italy}

\abstract
{}
{We present an analysis of the Lyman continuum (LyC) escape fraction of high redshift galaxies using an analytic post-processing approach.}
{We apply the model to $\sim 6 \times 10^5$ galaxies of the Illustris TNG50 simulation in the redshift range $z=5.2-20$.}
{Our study reveals a bimodal nature of LyC escape, which is associated either to (a) high metallicity ($10^{-3.5}<Z<10^{-2}$), low mass ($M_\star<10^7\mathrm{M}_\odot$) galaxies which due to their efficient cooling exhibiting extended star formation, with photons escaping primarily from the outer regions of the galactic plane ({\it ext}-mode), or (b)  low metallicity ($Z<10^{-3}$), moderately massive galaxies ($M_\star<10^8\mathrm{M}_\odot$) in which star-formation can only take place in small high density regions with localized LyC escape originating from these regions ({\it loc}-mode). While the {\it loc}-mode is present at all redshifts under investigation, the {\it ext}-mode becomes prominent in small galaxies at later cosmic times, once sufficient metal enrichment has occurred. Building on these findings, we develop an analytical fitting formula to determine the escape fraction of galaxies based on their stellar and gas mass, as well as redshift, providing a valuable subgrid modelling tool for future studies.}
{}

\maketitle

\section{Introduction}

The escape of Lyman continuum (LyC) photons, primarily emitted by young massive stars residing in galaxies, is a critical process driving the reionization of the Universe (see \citealt{dayal18} for a comprehensive review). To understand the history and structure of this cosmic event, it is crucial to gain insights into the LyC escape fraction. Indeed, by quantifying the fraction of LyC photons that successfully escape from galaxies and understanding the underlying physical mechanisms that facilitate or hinder this escape, we can obtain valuable information about the sources, mechanisms, and conditions contributing to the reionization process.

Due to the attenuation caused by neutral gas, the direct observation of LyC escape is limited to low redshifts, where typical escape fractions $\lesssim 0.05$ have been measured \citep[e.g.][]{vanzella10, nestor11, boutsia11, mostardi13, grazian16, mathee16, alavi20, pahl21}. These values are insufficient to meet the current constraints on the timing of reionization. However, some authors (e.g. \citealt{inoue06}) have suggested that in the redshift range $1 \leq z \leq 4$ the escape fraction increases with redshift, indicating that during the epoch of reionization, LyC escape might have been significantly higher.

To investigate LyC escape at higher redshifts, indirect indicators are employed, such as the ratio of the [OIII]5007Å/[OII]3727Å lines \citep[e.g.][]{nakajima14, tang21}. High values of this ratio  suggest that galaxies are density- rather than ionization-bound, i.e. they are more likely of being LyC leakers. Additional metal lines, such as MgII \citep[e.g.][]{chisholm20, xu16}, CII \citep[e.g.][]{heckman01, leitet11}, as well as SiII and SiIII \citep[e.g.][]{jones13, chisholm17}, can also be utilized to study the presence of neutral and ionized regions, providing further insights into the occurrence of LyC escape. 
Ly-$\alpha$ equivalent width \citep[e.g.][]{dijkstra14} and line separations \citep[e.g.][]{verhamme15} have also been employed as indicators of LyC escape. Indeed, a larger equivalent width indicates a higher rate of LyC production within the galaxy, as the Ly-$\alpha$ line primarily arises from hydrogen recombination. 
On the other hand, line separations are created by the scattering of Ly-$\alpha$ photons in neutral gas, so that a smaller line separation indicates lower amounts of intervening neutral gas and, consequently, a larger LyC escape fraction. 
Finally, it has been observed that the slope of the non-ionizing UV spectrum exhibits a strong correlation with LyC escape \citep{chisholm22}, with a redder UV slope indicating stronger dust attenuation, i.e. a significant absorption of LyC radiation. Therefore, galaxies with steeper UV slopes tend to have lower LyC escape fractions. 

Theoretical investigations have endeavored to model the escape of LyC radiation from galaxies. These studies employ methodologies such as Monte Carlo radiation transfer (MCRT) in post-processing \citep[e.g.][]{paardekooper15, ma16, ma20, kostyuk23}, as well as hydrodynamic RT simulations \citep[e.g.][]{kimm14, ocvirk16, rosdahl18, lewis20, ocvirk21, rosdahl22, yeh22}. However, determining the escape fraction poses challenges due to degeneracies between star formation efficiency and escape fraction, as well as the large dynamic range involved, requiring additional assumptions about the subgrid configuration. As a consequence, various studies report average values of the escape fraction that can vary up two orders of magnitude, in addition to finding different dependencies of the escape fraction on galactic properties, such as stellar and gas mass, although most agree that the escape fraction decreases with increasing mass for the highest stellar masses.

An alternative approach to understanding LyC escape is currently being developed by Ferrara et al. (in prep., F24). This involves constructing an analytic model that can identify LyC leakage from a galaxy based on a few simple assumptions. The model represents the galaxy as a thin slab, with dusty gas surrounding the galactic plane acting as the absorber of LyC radiation. With these assumptions, the model aims to provide insight into the mechanisms and conditions governing LyC escape, offering a streamlined framework for studying this phenomenon in galaxies. Due to the analytical nature of the model and the broad spectrum of physical scales involved in LyC escape, the objective is not to determine precise values for the escape fraction, but rather to acquire a qualitative insight into the physical characteristics of galaxies that leak LyC radiation.
In this study, we employ the aforementioned physical model to investigate the characteristics of LyC radiation in galaxies from the TNG50 cosmological simulation. The aim of this work is to provide an introduction into the capabilities of a simplified semianalytic model to capture qualitative behaviour of LyC escape at high redshifts. We aim to further expand the complexity and reliability of the model in future works.
%Our approach involves evaluating LyC escape on a regular grid, enabling us to identify regions within a galaxy where LyC leakage occurs. By applying this methodology, we aim to gain insights into the spatial distribution and properties of LyC escaping from galaxies.

We structure this work as follows: in section \ref{sec:semi_method} we present our methodology for modeling LyC photon production and escape in TNG50 galaxies. In section \ref{sec:semi_results} we discuss the geometry of LyC escape as well as its correlation with galactic properties. We subsequently introduce a fitting formula for the LyC escape fraction and analyze its accuracy in section \ref{sec:semi_fitting}. We finally summarize our results and conclude in section \ref{sec:semi_conclusions}.

\section{Method} \label{sec:semi_method} 
Here we present our approach for calculating the LyC escape fraction of TNG50 galaxies using the semianalytic model developed in F23. We first introduce the TNG simulations, followed by a description of the physically motivated model adopted to evaluate the escape fraction, and finally, we present its application to galaxies extracted from TNG50.

\subsection{The IllustrisTNG simulation}

In this work we model the LyC escape fractions of galaxies taken from the Illustris TNG50 simulation \citep{pillepich18b, springel18, naiman18, nelson18a, marinacci18}, which is one of three large scale cosmological simulations, together with TNG100 and TNG300, sharing the same physical modelling. TNG50 simulates the evolution of structures in a cosmological volume of $(51.7\mathrm{cMpc})^3$ 
\citep{nelson19a, pillepich19}. The initial conditions were generated with the N-GenIC code \citep{springel05} at $z=127$, adopting cosmological parameters consistent with the measurements by the Planck satellite \citep{planck16}, i.e. $\Omega_\mathrm{m}=0.3089$, $\Omega_\mathrm{b}=0.0486$, $\Omega_\Lambda=0.6911$, $h=0.6774$, $\sigma_8=0.8159$ and $n_s=0.9667$. The simulation evolves down to $z=0$, and the properties of galaxies are validated against observational data, providing confidence that the high-redshift galaxies analysed in this study are realistic progenitors of the galaxy populations observed today.

The underlying physical model utilized in this study is implemented using the {\textsc AREPO} code \citep{springel10}, which employs the equations of ideal magnetohydrodynamics to simulate the interaction of baryonic matter. The simulation is performed on a Voronoi grid, enabling a detailed representation of the spatial distribution and properties of the simulated system. The galaxy formation model employed in this work builds upon the framework developed by \cite{weinberger17} and \cite{pillepich18a}. In this model, the gas particles contained within Voronoi cells with a hydrogen number density $n_\mathrm{H}> 0.1 \mathrm{cm}^{-3}$, determined based on the Kennicutt-Schmidt relation \citep{springel03}, are stochastically converted into individual stellar particles. Each stellar particle represents a distinct stellar population and follows a Chabrier initial mass function (IMF; \citealt{chabrier03}). To account for subgrid processes and the role of pressure support in star-forming gas, the \cite{springel03} model for the interstellar medium (ISM) is incorporated. Feedback from supernovae is modeled through the implementation of decoupled kinetic wind particles \citep{pillepich18a}.

The initial dark matter and gas particles mass is $4.5 \times 10^5\mathrm{M}_\odot$ and $8.5 \times 10^4 \mathrm{M}_\odot$, respectively. However, the refinement and de-refinement of Voronoi cells leads to variation of these values within a factor of two. The spatial resolution varies strongly from the inner part of the galaxies to the IGM. The mean size of star forming gas cells, which are of most relevance for our study, is 144 physical parsecs at $z=6$, with 16-84th percentiles ranging from 96 to 184~ppc. Whenever a stellar particle is formed, it inherits the mass of the baryonic gas cells it formed in, and its mass decreases as the stellar population evolves with time. 

The galaxies we investigate in this study are identified as TNG50 subhalos. These are found using the SubFind algorithm \citep{springel01}, which employs the friends-of-friends approach \citep{davis85} to detect distinct structures within the simulation. Subhalos are defined based on a minimum particle threshold, with a minimum subhalo mass of approximately $10^6\mathrm{M}_\odot$. At the center of galaxies surpassing a mass threshold of $10^{10.8}\mathrm{M}_\odot$, black hole (BH) seeds with an initial mass of $8\times 10^5 h^{-1} \mathrm{M}_\odot$ are positioned. The growth of the seeds occurs through either BH-BH mergers or gas accretion, following the Bondi formalism with the Eddington rate setting the accretion limit. To account for the feedback effects of black holes, the simulation incorporates various mechanisms such as thermal heating, radiative influences, and kinetic wind \citep{weinberger17}. Moreover, the simulation tracks the metal enrichment process, including the individual abundances of elements such as C, N, O, Ne, Mg, Si, Fe, and Eu, in addition to modeling the content of H and He gas. While radiation transfer is not modelled explicitly, a spatially uniform UV-background (UVB; from \citealt{fg09}) is turned on at $z<6$. 

\subsection{Physical model of LyC escape}

The model underlying our calculations of the escape fraction was developed in F23. Here, the galaxy is considered a plane parallel slab (as in \citealt{ferrara19}), with stars populating the galactic plane, while the gas is distributed above and below it with a constant density up to a scale height of $H$, resulting in a column density $N_0$. This is a simplification of the typically more complex geometry of the gas distribution. However, it is a reasonable approximation given a gas distribution that is concentrated around the immediate surrounding of the source with only a few cells significantly contributing to gas absorption. A more realistic picture would involve the resolution of the two-state ISM which is not included in the TNG50 model.

LyC photons emitted by stars are absorbed by both gas and dust, and they can escape the galaxy if the ionizing flux is large enough to form an ionized channel in the gas. Although lower mass galaxies at high redshift typically lack a disc-like structure, the fundamental physical assumption underpinning this model is that the scale height of the star forming region is notably smaller than the height of the gas distribution. This assumption is further examined in appendix \ref{app:sf-scale}, where we show that statistically it holds over the full range of galactic masses we examined.

We define the ionizing photon flux $F_i$ (in units of $\mathrm{cm}^{-2}\mathrm{s}^{-1}$) as:
\begin{equation}
    F_i = \int_{\nu_L}^{\infty} \frac{F(\nu)}{h_P\nu} \mathrm{d}\nu, 
    \label{eq:ion flux}
\end{equation}
where $\nu$ is the frequency, $\nu_L = 3.2 \times 10^{15}$Hz is the frequency corresponding to the H ionization threshold 13.6~eV, $F(\nu)$ is the ionizing energy flux, and $h_P$ is the Planck constant. 

As in \cite{ferrara19}, the ionizing energy flux is expressed through a power law
\begin{equation}
    F(\nu) = F_L \left( \frac{\nu}{\nu_L} \right)^{-\beta},
    \label{eq:luminosity scaling}
\end{equation}
where $F_L$ is the energy flux at $\nu_L$, while $\beta$ is usually assumed to be in the range $4-5$ for PopII stellar populations \citep{leitherer99}. In this work we use $\beta=4$. Therefore, eq.\ref{eq:ion flux} becomes $F_i = F_L/(\beta h_P)$. Here, we obtain $F_L$ through its connection to the SFR. For this purpose, we use the conversion between UV luminosity $F_{1500}$ and the surface density of star formation $\Sigma_\mathrm{SFR}$ employed by the REBELS collaboration \citep{dayal22}, i.e. $F_{1500} = \Sigma_\mathrm{SFR} \kappa^{-1}$. For a Chabrier IMF, $\kappa =7.1\times 10^{-29} \mathrm{M}_\odot \mathrm{yr}^{-1}\mathrm{erg}^{-1}\mathrm{s}\mathrm{Hz}$ and  $F_{1500} \approx F_L$. Hence, the ionizing flux can be calculated as
\begin{equation}
    F_i = \frac{\Sigma_\mathrm{SFR}}{\beta h_P \kappa}. \label{eq:Fi}
\end{equation}

Given the gas density, the photon flux can be related to the photon-to-gas density ratio $U$ through $U = F_i/nc$, with $c$ being the speed of light. We can further relate the flux to the Str{\"o}mgren depth $l_S$ (i.e. the maximum distance that can be ionized by the flux) through $l_S = {F_i}/(n^2 \alpha_B)$,
with $\alpha_B$ being case B recombination parameter. We can then define the maximum column density that can be ionized by the flux:
\begin{equation}
    N_S = nl_S = \frac{Uc}{\alpha_B}, \label{eq:ns}
\end{equation}
and the optical depth encountered by LyC photons through the hydrogen gas:
\begin{equation}
    \tau_\mathrm{HI} = \frac{N_0}{N_S}.
\end{equation}

The ionizing flux is additionally absorbed by the dust content in the gas, which has an optical depth to LyC photons of:
\begin{equation}
    \tau_d = \left(\frac{A_L}{A_V} \right) \sigma_{d,V} N_0. \label{eq:tau_d}
\end{equation}
Here $A_L/A_V=4.83$ is the LyC to V-band attenuation ratio assuming a visual extinction to reddening ratio of $R_V = 3.1$\footnote{It should be noted that this ratio is based on Milky Way extinction curves and could be sensitive to redshift evolution.} and $\sigma_{d,V}= 4.8 \times 10^{-22} \mathcal{D} \mathrm{cm}^2$ is the  dust cross section per H atom in the V-band \citep{weingartner01}. 
Assuming a linear relation between metallicity $Z$ and dust, the dust-to-gas ratio normalized to the galactic value $\mathcal{D}$ can be written as $\mathcal{D}= Z/Z_\odot$, with $Z_\odot=0.0134$ being the solar metallicity \citep{asplund09}.

%The dust cross section per H atom in the V band is given by
%\begin{equation}
%    \sigma_{d,V} = 4.8 \times 10^{-22} \mathcal{D} \mathrm{cm}^2.
%\end{equation}
%Assuming a linear relation between metallicity $Z$ and dust, the dust-to-gas ratio normalized to the galactic value $\mathcal{D}$ can be written as $\mathcal{D}= Z/Z_\odot$,
%where $\mathcal{D}$ represents the dust-to-gas ratio normalized to the galactic value. Assuming a linear relation between the metallicity $Z$ and dust, the dust content can be written as 
%\begin{equation}
%\mathcal{D}= \frac{Z}{Z_\odot},
%\end{equation}
%with $Z_\odot$ solar metallicity
%{\bc which one are you using? Asplund?}{ \ik Weingartner and Draine 2001, I think. It is only later cited in Andrea's paper but the 4.8 value seems to agree with fig 19 of that paper.}
%{\bc I meant which value of the solar metallicity you assume...from Asplund?}
%The dust optical depth to LyC photons
%\begin{equation}
%    \tau_d = \left(\frac{A_L}{A_V} \right) \sigma_{d,V} N_0.
%\end{equation}
%Here the LyC to V-band attenuation ratio was used which according to \cite{weingartner01} and assuming an $R_V = 3.1$ is given by $A_L/A_V = 4.83$. It should be noted that this ratio is based on Milky Way extinction curves and could be sensitive to redshift evolution. 
We can thus determine the gas column density at which the dust optical depth reaches unity as:
\begin{equation}
    N_\mathrm{d} = 4.3 \times 10^{20} \mathcal{D}^{-1} \mathrm{cm}^{-2}. \label{eq:Nd}
\end{equation}
This simplifies eq.~\ref{eq:tau_d} to $\tau_d = N_0/N_\mathrm{d}$.

The LyC escape fraction is defined as the fraction of photons not absorbed by the optical depth, i.e. $f_\mathrm{esc} = e^{-\tau_\mathrm{tot}}$, where $\tau_\mathrm{tot}=\tau_\mathrm{HI}+\tau_{d}$ is the total optical depth of the gas in the galaxy. Hence, we can write the escape fraction as\footnote{Note that F23 take into account also a reduction in $N_0$ due to radiation pressure pushing out part of the gas. However, as radiation pressure is not explicitly included in the TNG50 simulations, we neglect this effect and adopt the above simplified model.}:
\begin{equation}
    f_\mathrm{esc} = \exp \left[-N_0\left(\frac{1}{N_S}+ \frac{1}{N_\mathrm{d}} \right) \right]. \label{eq:escape fraction}
\end{equation}

\subsection{Escape fraction of TNG50 galaxies}

\begin{figure}
    \centering
    \includegraphics[width=0.49\textwidth]{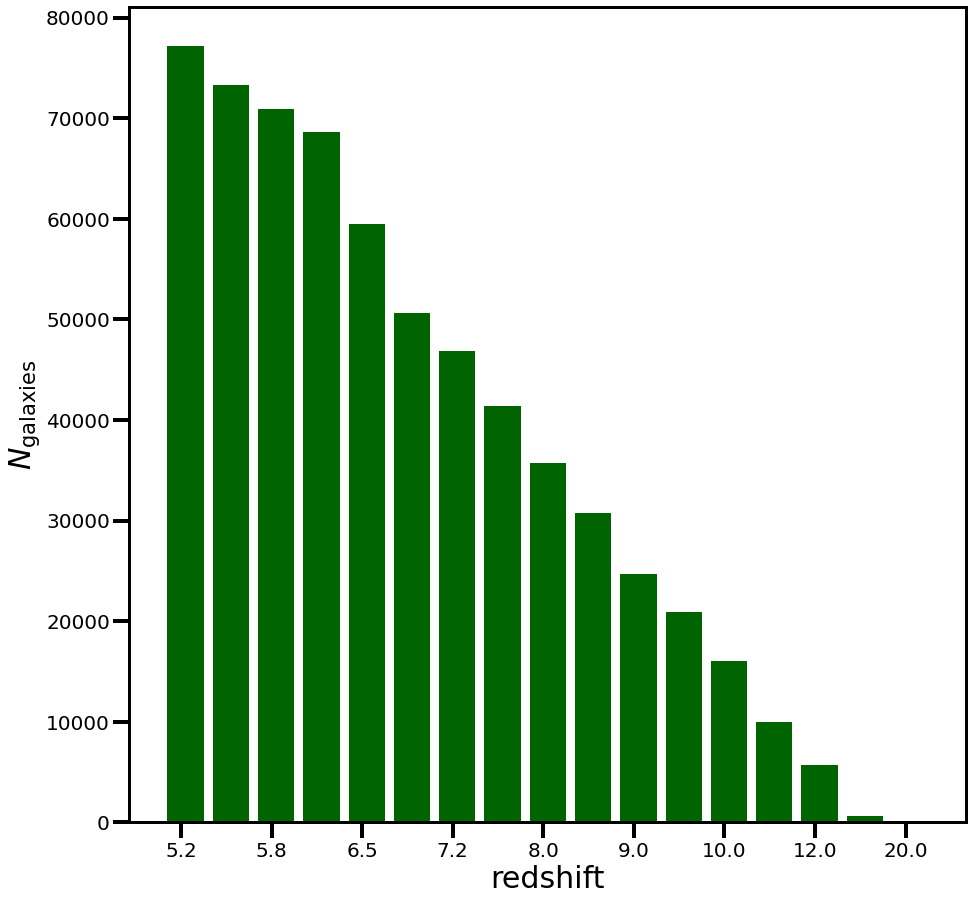}
    \includegraphics[width=0.49\textwidth]{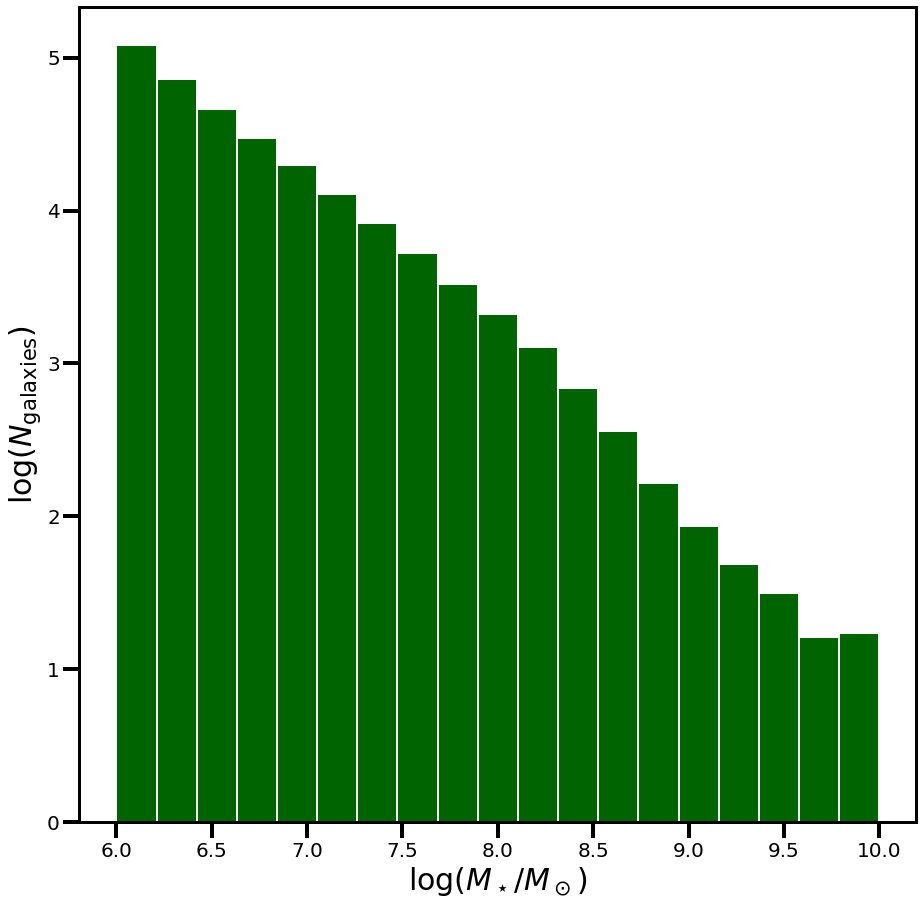}
    
    \caption{Number of galaxies processed as a function of redshift \textit{(top)} and as a function of stellar mass \textit{(bottom)}}.
    \label{fig:counts}
\end{figure}

In this work, we analyse the escape fraction of $\approx 600,000$ galaxies extracted from 17 TNG50 snapshots in the redshift range $5.2 \leq z \leq 20$. 
To ensure sufficient resolution, we post-process only galaxies with a minimum stellar and gas mass of $10^{5.5}\mathrm{M}_\odot$ and $10^{6}\mathrm{M}_\odot$, respectively, located within twice their stellar half mass radius. This ensures that the smallest galaxies are resolved with more than 10 gas particles. This results in $77498$ galaxies at $z=5.2$ and only $6$ at $z=20$ (see fig.~\ref{fig:counts} top). As the TNG50 snapshots are not equally spaced in redshift and a much larger number of galaxies have formed at later times, the lower redshift range is sampled with significantly more galaxies. Additionally, given their much larger abundance the sample is highly biased towards low mass galaxies. We sample a factor of $10^4$ more galaxies with $M_\star \approx 10^6M_\odot$ as compared to galaxies with $M_\star \approx 10^{10}M_\odot$ (see fig.~\ref{fig:counts} bottom).

To obtain the LyC escape fraction for each galaxy we proceed as follows. First, we define the galactic plane as the surface spanned by the first two principle axes of the galaxy. Then, we cut out a box with a side length of four times the stellar half mass radius, $r_\mathrm{HMR}$. The gas particles contained in that box are distributed on a regular Cartesian grid with $100^3$ cells (see appendix \ref{app:convergence_test} for a convergence study), ensuring that both the gas distribution and the star forming regions are well resolved.
We define the gas scale height, $H$, as the height above and below the galactic plane, containing $\approx 63\%$ ($1/e$) of the gas mass. 

We then project quantities relevant to the calculation of the escape fraction onto the galactic plane. More specifically, the gas mass per unit area in each 2D grid cell $\{i,j\}$ is given by:
\begin{equation}
    \Sigma_{m_{i,j}} = \frac{\sum\limits_k m_{\mathrm{gas},ijk}}{A_\mathrm{cell}},
\end{equation}
with $m_{\mathrm{gas},ijk}$ being the gas mass contained in a given grid cell, and $A_\mathrm{cell}$ area of the cell. 
$\Sigma_{\rm SFR_{i,j}}$ is calculated in the same way, by substituting $m_{\mathrm{gas},ijk}$ with the instantaneous SFR in each cell SFR$_{ijk}$. $\Sigma_{m_{i,j}}$ can then be related to the gas column density in the cell through $N_{0,ij}=\Sigma_{m_{i,j}}/2\mu$, with $\mu$ being the mean molecular mass assumed to be the mass of a hydrogen atom in our simplified model, and the factor $2$ accounts for the symmetry of the gas distribution below and above the galaxy.
Finally, the metallicity of a grid cell $Z_{i,j}$ is given by the mass weighted mean in a given column:
\begin{equation}
    Z_{i,j} = \frac{\sum\limits_k Z_{ijk}m_{\mathrm{gas},ijk}}{\sum\limits_k m_{\mathrm{gas},ijk}}.
\end{equation}
This value can be related to the value of $N_\mathrm{d}$ in that cell through eq.~\ref{eq:Nd}.

To obtain the escape fraction from a cell, we compute $N_0$, $N_S$ and $N_\mathrm{d}$ individually for each element of the 2D grid and evaluate $f_\mathrm{esc,cell}$ using eq.~\ref{eq:escape fraction}. 
The escape fraction of the galaxy, $f_\mathrm{esc}$, is then given by:
\begin{equation}
    f_\mathrm{esc} = \frac{\sum\limits_\mathrm{cell} F_{i,\mathrm{cell}}f_\mathrm{esc,cell}}{\sum\limits_\mathrm{cell} F_{i,\mathrm{cell}}},
\end{equation}
with $F_{i,\mathrm{cell}}$ being the ionizing flux in each grid cell obtained using eq.~\ref{eq:Fi} with $\Sigma_{\rm SFR_{i,j}}$. 

It should be noted that neither the analytic method for the calculation of the escape fraction nor the hydrodynamic model used to simulate star formation are fully taking into account the multi-phase structure of the ISM. While we anticipate that an improved subgrid modeling of the complex nature of the ISM would lead to quantitative differences in some of our results, we remain confident that the qualitative conclusions will remain unchanged. This is supported by the study conducted in \cite{kostyuk23}, where we demonstrated that the inclusion of a subgrid escape fraction, while affecting the absolute value of $f_{\rm esc}$, did not alter its dependence on galactic properties. Furthermore, we also note that, as of now, no hydrodynamical simulation is able to cover all the relevant scales involved, i.e. all studies of escape fraction are necessarily resolution limited.

\section{Results} \label{sec:semi_results}
\subsection{Dependence on stellar mass and redshift \label{sec:general_res}}

\begin{figure}
    \centering
    \includegraphics[width=0.45\textwidth]
    {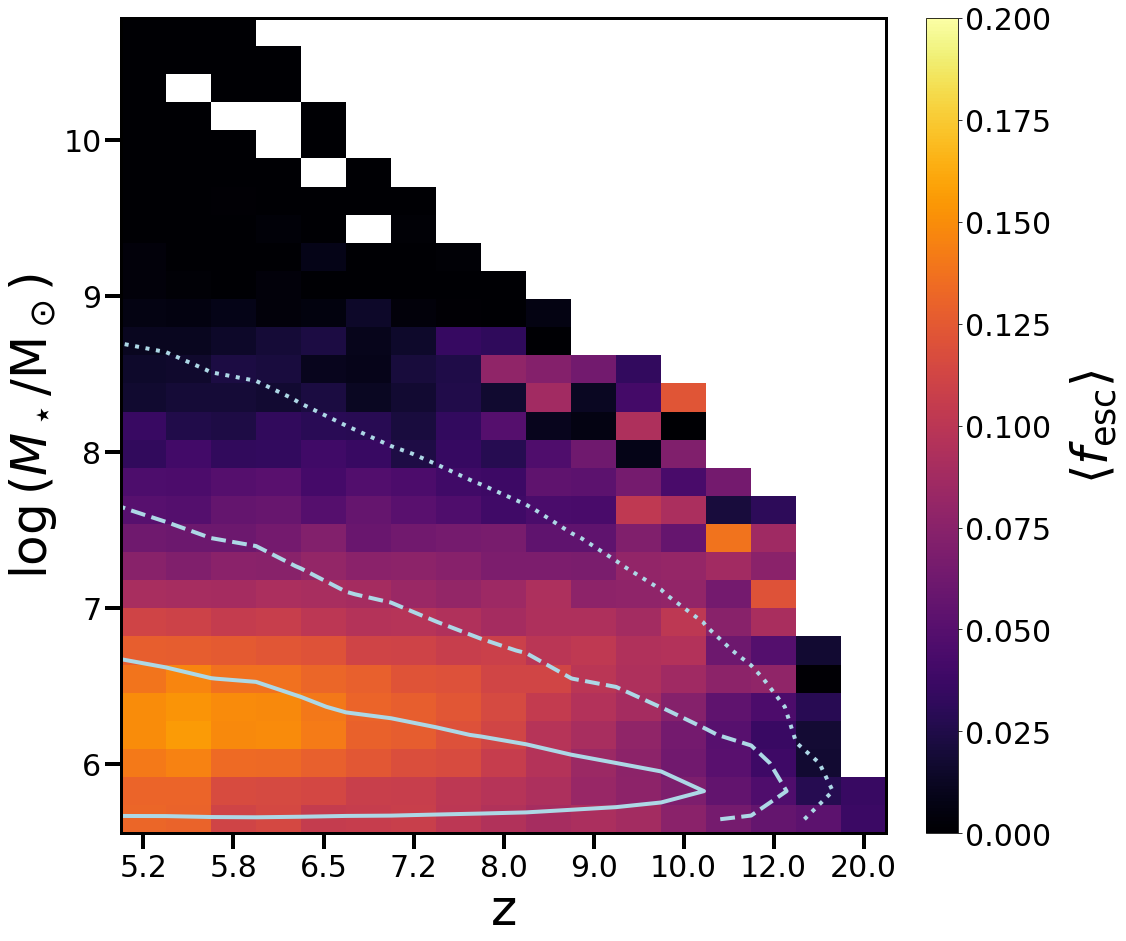}
    % \hspace{0.5cm}
    \includegraphics[width=0.45\textwidth]
    {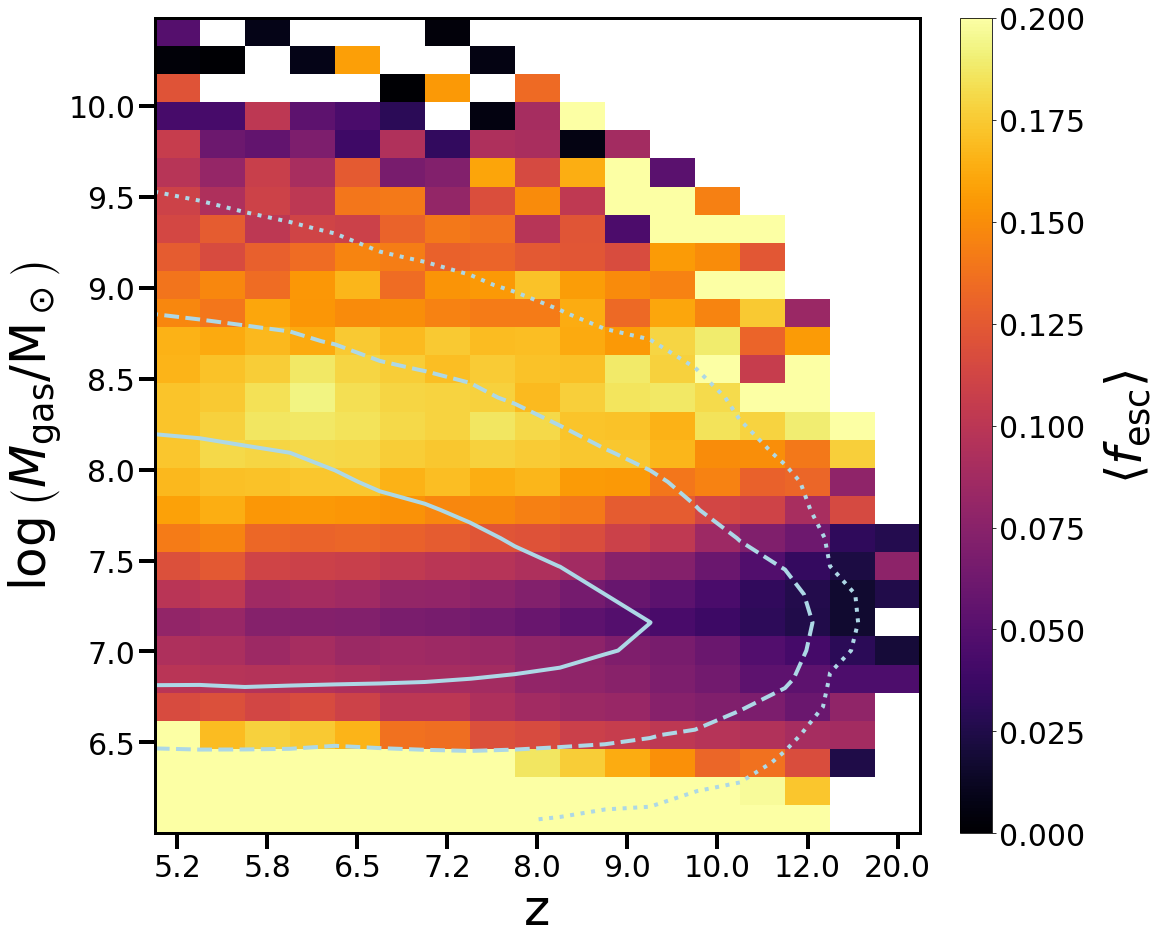}
    \caption{Average LyC escape fraction of all galaxies in a given bin (color scale) as a function of redshift and galactic stellar (top) as well as gas (bottom) mass. The white contours denote the 1  (solid), 2 (dashed) and 3 (dotted) sigma distribution of galaxy counts.
    }
    \label{fig:general histograms}
\end{figure}

\begin{figure}
    \centering
    \includegraphics[width=0.45\textwidth]
    {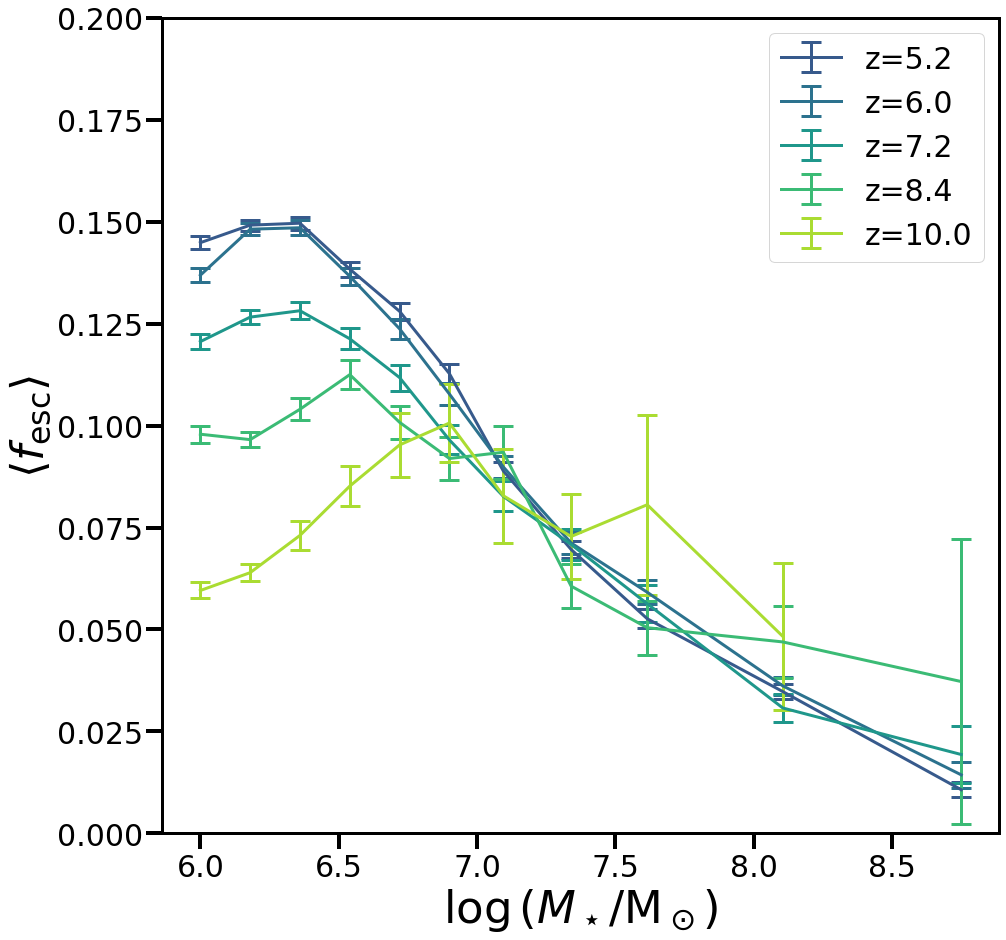}
    % \hspace{0.5cm}
    \includegraphics[width=0.45\textwidth]
    {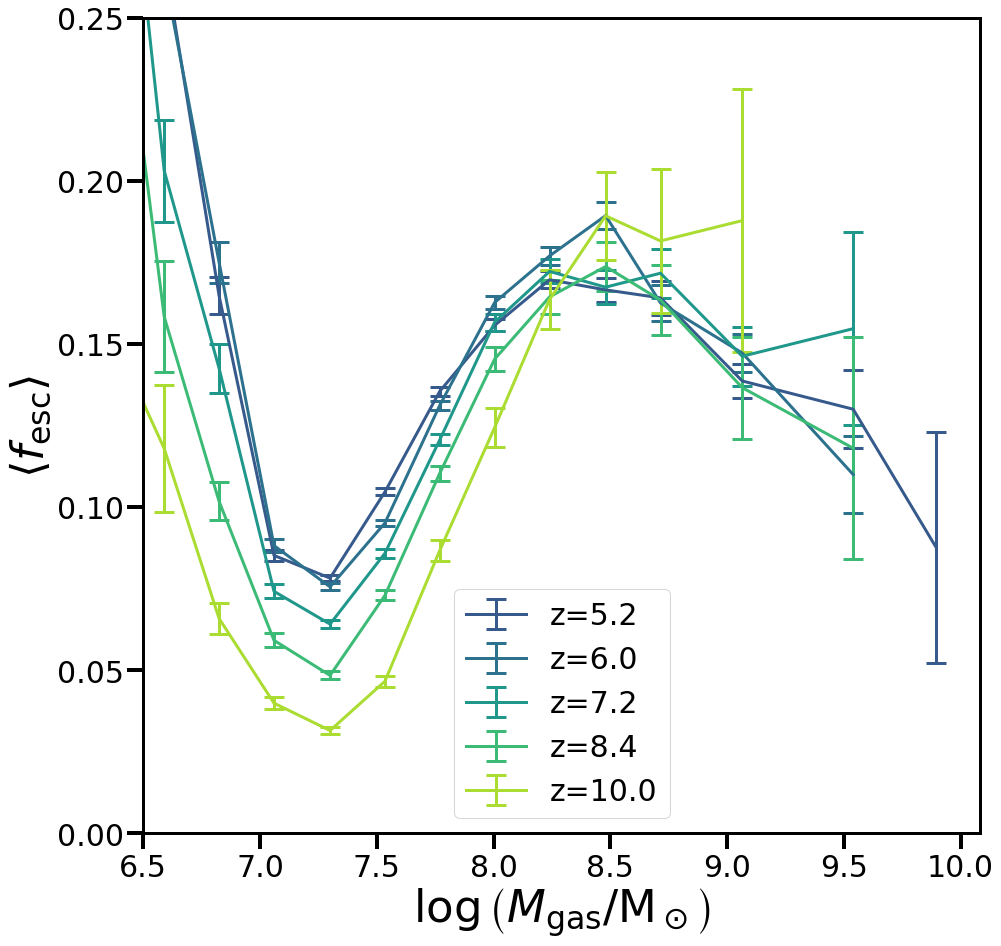}
    \caption{Average LyC escape fraction as a function of stellar mass (top) and gas mass (bottom) for a sample of redshifts. The error bars denote the statistical error of the mean. }
    \label{fig:line plots}
\end{figure}

In fig.~\ref{fig:general histograms} we investigate how the LyC escape evolves as a function of the galactic stellar and gas mass. For this purpose, we classify galaxies based on their redshift, as well as their stellar and gas mass. We then represent the averaged escape fraction $\langle f_\mathrm{esc} \rangle$ of galaxies within each specific bin using a color scale.
We observe a strong correlation between stellar mass and the average escape fraction, with $\langle f_\mathrm{esc} \rangle$ steadily increasing with $M_\star$ up to $M_\star \approx 10^{6.5}\mathrm{M}_\odot$ ($10^7\mathrm{M}_\odot$) for $z=5.2$ (10). Beyond that, $\langle f_\mathrm{esc} \rangle$ decreases and becomes essentially zero for $M_\star>10^8$~M$_\odot$. As redshift increases, the peak of the escape fraction shifts towards higher stellar masses. This phenomenon is more clearly discernible in fig.~\ref{fig:line plots}, which will be discussed later.

A larger scatter is observed in the dependence of the escape fraction on the galactic gas mass, which emerges as a result of the gas mass exhibiting a weaker correlation (in comparison to the stellar mass) with the metallicity. A higher metallicity results in an increase in dust content and hence absorption, as well as in more efficient cooling and thus denser gas, with both factors being strongly anti correlated with the escape fraction, as will be further discussed in fig.~\ref{fig:properties escape modes}.
The generally higher $\langle f_\mathrm{esc} \rangle$ as compared to the top panel, is explained by the majority of the high-escape bins being located outside of the regions containing most of the galaxies, so that the bins are typically sampled with fewer galaxies.
We observe that the escape fraction exhibits one maximum at $M_\mathrm{gas}=10^8\mathrm{M}_\odot$ at $z=5.2$, which increases with redshift as in the top panel, reaching $M_\mathrm{gas}=10^{8.5}\mathrm{M}_\odot$ at $z=10$.
A second peak observed for galaxies with $M_{\rm gas}<10^{6.5}\mathrm{M}_\odot$, shows instead an opposite redshift evolution. This peak can be attributed to a secondary mode of LyC escape correlated to extensive star formation regions and elevated metallicity levels. We thoroughly investigate this phenomenon in fig.~\ref{fig:Escape_columns_height}.

In fig.~\ref{fig:line plots} we show the averaged escape fraction as a function of stellar and gas mass for a sample of redshifts. 
Here we see more clearly how the peak of $\langle f_\mathrm{esc} \rangle$ shifts towards higher stellar masses with increasing redshift, while at the same time its value decreases. For $M_\star \geq 10^7\mathrm{M}_\odot$, though, $f_\mathrm{esc}$ converges to similar values at all redshifts. It should be noted that the high redshift results are less reliable due to the smaller sample of galaxies (see fig.~\ref{fig:counts}). 

As already seen in fig.~\ref{fig:general histograms}, the escape fraction decreases with increasing gas mass, reaching a minimum at $M_{\rm gas} \approx 10^{7.3}\mathrm{M}_\odot$ before increasing again for higher gas masses. Note that this initial decrease is particularly strong for lower redshifts, reflecting the trend in fig.~\ref{fig:general histograms}. 
Additionally, the lower panel shows a smaller difference in the amplitude of the peak across the different redshifts. The increase in the escape fraction of galaxies with the smallest gas masses is likely to reflect the population of galaxies that have some star-formation but almost no gas obscuration.

\subsection{Escape modes \label{sec:semi_modes}}

Next, we investigate in more detail the correlation of $f_\mathrm{esc}$ with various galactic properties. 

\begin{figure}
    \includegraphics[width=0.49\textwidth]{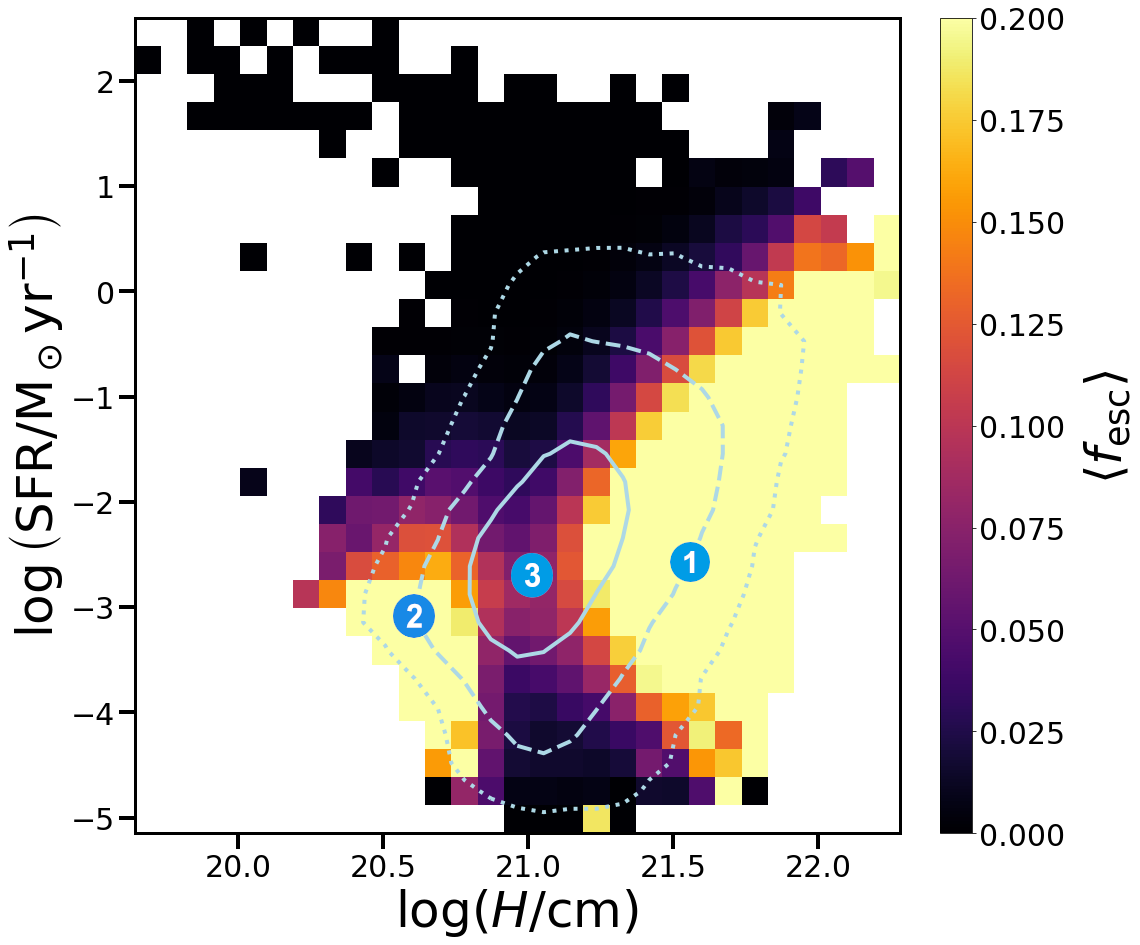}
    \includegraphics[width=0.49\textwidth]{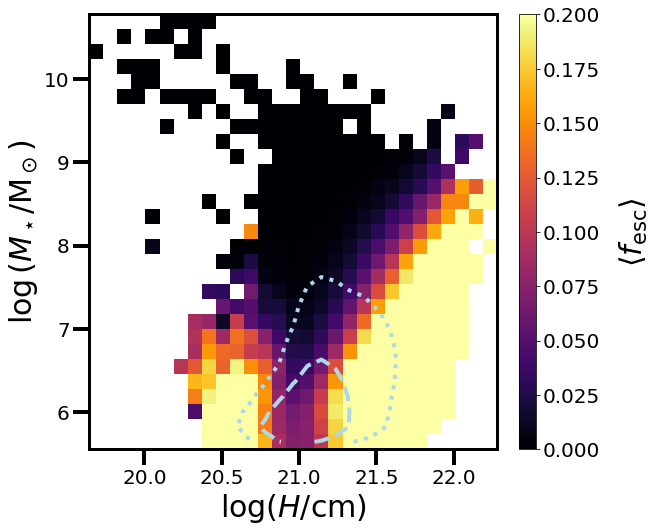}
    
    % \begin{subfigure}{0.5\textwidth}
    % \caption*{Sample galaxies from the localized escape mode (region 1)}
    % \includegraphics[width=\textwidth]{figures/high_h_sample.png} 
    % \end{subfigure}
    % \begin{subfigure}{0.5\textwidth}
    % \caption*{Sample galaxies from the extended escape mode (region 2)}
    % \includegraphics[width=\textwidth]{figures/low_h_sample.png}
    % \end{subfigure}
    % \begin{subfigure}{0.5\textwidth}
    % \caption*{Sample galaxies from the boundary region (region 3)}
    % \includegraphics[width=\textwidth]{figures/center_fesc.png}
    % \end{subfigure}
    \caption{Average escape fraction as a function of scale height and star formation rate \textit{(top)} as well as a function of scale height and galactic stellar mass \textit{(bottom)}. The contours represent the distribution of galaxies as described in fig. \ref{fig:general histograms}. The numbers label the areas that we identified as the localized escape mode (\textit{loc}-mode, number 1), the extended escape mode (\textit{ext}-mode, number 2) and the LyC dark mode (number 3). 
    % Bottom panel: Projected maps of the escape fraction distribution from individual grid cells (box size $4$ stellar half mass radii) for three example galaxies from each of the regions indicated by numbers. {\bc what is the extent of the maps? are they slices or projected? what's the fesc of the single galaxies? I'm asking these questions because if you e.g. compare the first galaxy of the localized and boundary region, the latter seem to have a much higher fesc, so the maps are deceiving}
    % {\ik I added 'projected' to the text. Yes indeed to see that the escape fractions are similar one has to weigh them by the ionizing flux maps in fig. 5. I originally had the total escape fraction in the label of the plot but that made it appear crowded. Here I only wanted to show the escape channels qualitatively. Should I add the absolute values in the titles again?} {\bc what is 4SHMR??? and yes, you should definitely write the escape fraction again, you can write it in the caption if the figures get too crwoded}{\ik changed to stellar half mass radii... I do introduce SHMR in the methodology section but if someone just scans the plots it might indeed be confusing.}
    }
    \label{fig:Escape_columns_height}
\end{figure}

Initially, we examine the impact of gas distribution within the galaxy on the escape of LyC radiation. To accomplish this, in fig.~\ref{fig:Escape_columns_height} we show the average escape fraction as a function of scale height and the star formation rate as well as the scale height and galactic stellar mass, which serves as a measure of how closely the gas is distributed in relation to the galactic plane.
We observe the presence of two distinct modes, denoted as localized escape mode (\textit{loc}-mode, corresponding to the area labeled with the number 1 in the top plot) and the extended escape mode (\textit{ext}-mode, number 2), both exhibiting $\langle f_\mathrm{esc} \rangle > 0.2$.\footnote{Note that the while the color scale saturates the average escape fractions within a given mode keeps increasing following the pattern indicated by the color gradients.} The \textit{loc}-mode is characterised by higher scale heights ($H > 10^{21}$ cm) and a wider range in SFR.
%Additionally, a power-law trend can be observed at higher SFR values, delineating LyC leaking and non-leaking galaxies, which reverses its direction at lower SFRs. 
The \textit{ext}-mode, on the other hand, exhibits smaller scale heights ($H<10^{21}$cm), and is limited to relatively low SFRs.
%with an upper bound dictated by a similar trend to the localized escape mode. However, we do not observe any lower bound for this mode.
The two regions are separated by a region with small LyC escape (number 3) which we will refer to as the LyC dark mode, where the average escape fraction is lower. This mode encompasses the majority of galaxies, as evidenced by the contour lines. In the lower plot we see that the two mode structure of the escape fraction can also be identified as a function of stellar mass. However unlike with SFR there is no lower stellar mass bound for the escape mode.

\begin{figure*}
    \centering
    \includegraphics[width=0.48\linewidth]{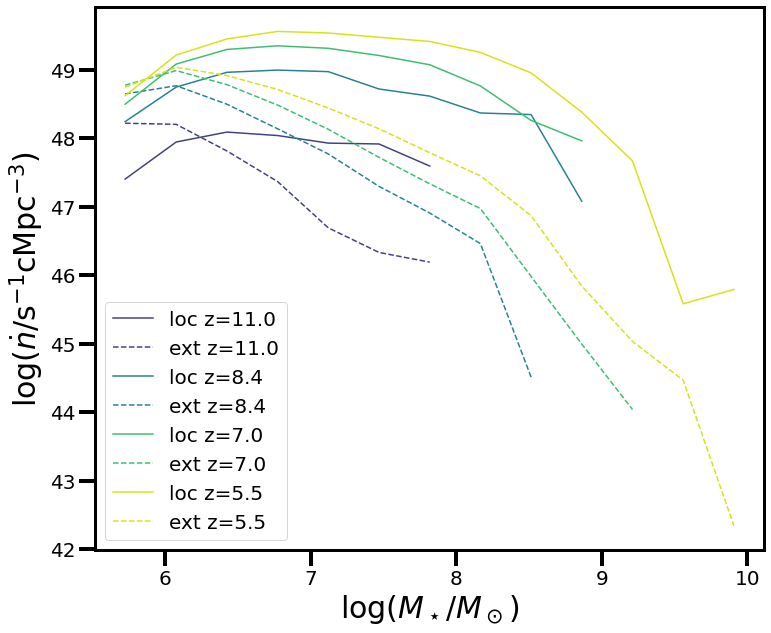}
    \includegraphics[width=0.48\linewidth]{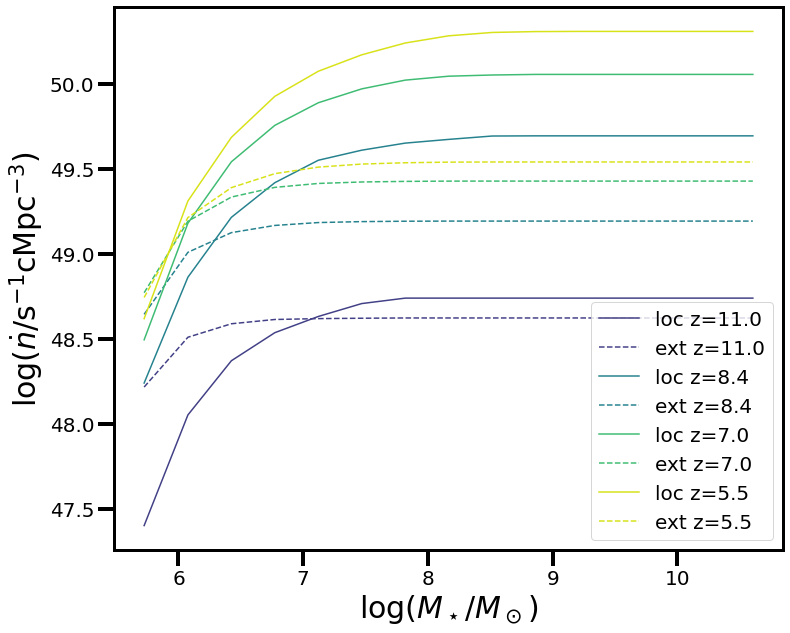}
    \caption{Comparison of the escape density of ionizing photons from the \textit{loc}-mode (solid) and \textit{ext}-mode (dashed) as a function of stellar mass (left) and well as the cumulative escape (right).}
    \label{fig:em_densities}
\end{figure*}

In fig.~\ref{fig:em_densities} we investigate the contributions of these two modes to the density of ionizing photon escape at various redshifts. We see that the \textit{ext}-mode dominates the LyC escape for low mass galaxies $M_\star \approx 10^6M_\odot$ while in all other mass bins the \text{loc}-mode is the dominant one. As small galaxies dominate the galactic population at high redshifts LyC leakage via \textit{ext}-mode contributes a comparable amount to the escaping ionizing photon budget as the \textit{loc}-mode at early times(see right-hand plot). However, at lower redshifts the \textit{ext}-mode quickly becomes subdominant contributing only $\approx 20\%$ of the ionizing photons at $z=5.5$.

% IK: I am now adding the paragraph about ndot here or would you suggest adding a separate paragraph

\begin{figure*}
    \centering
    \begin{subfigure}[t]{0.45\textwidth}
    \caption*{\textit{loc}-mode}
    \includegraphics[width=0.97\textwidth]{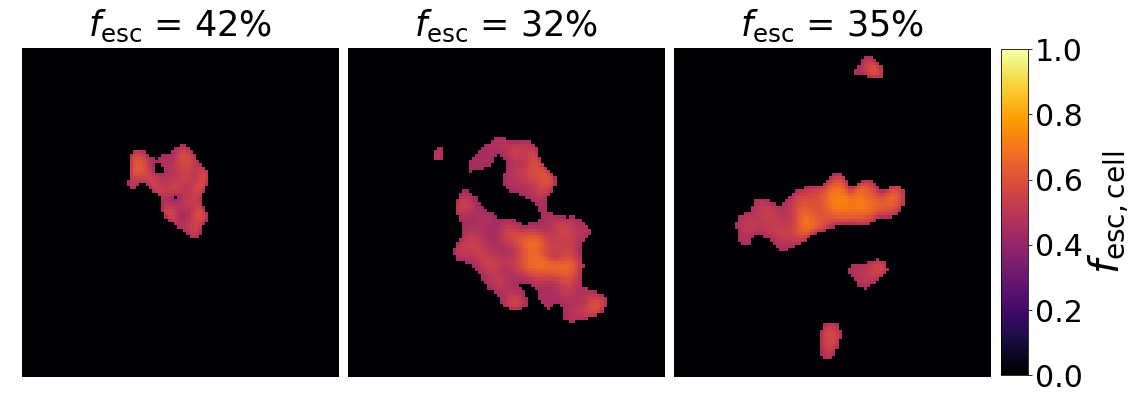} 
    \includegraphics[width=0.95\textwidth]{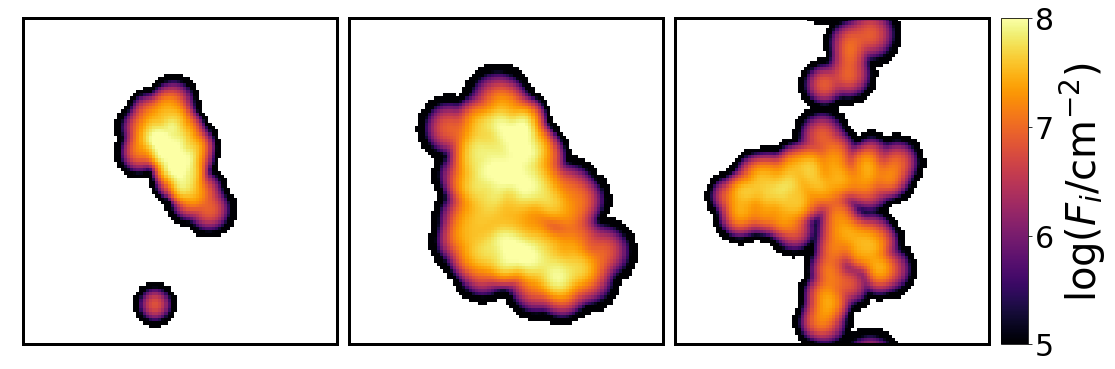} \\
    % \hspace{0.2cm}
    % \hspace{0.cm}
    \includegraphics[width=0.98\textwidth]{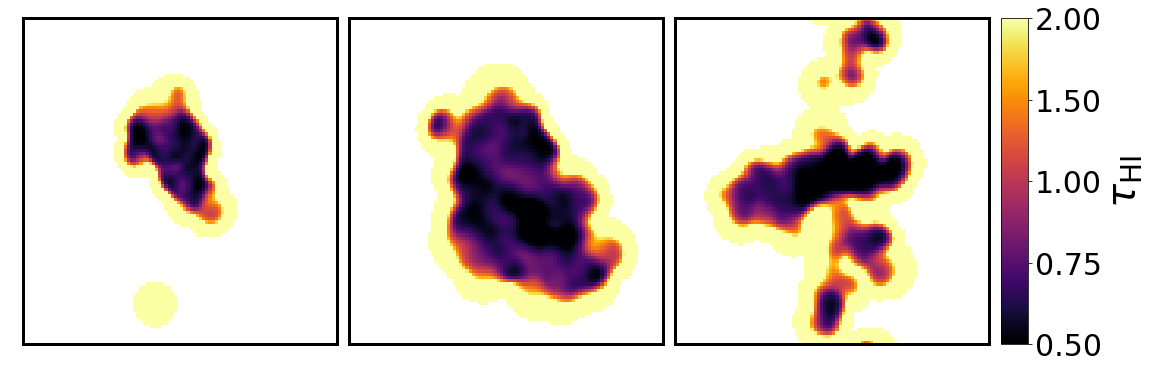} 
    \end{subfigure}
\qquad
    \begin{subfigure}[t]{0.45\textwidth}
    \caption*{\textit{ext}-mode}
    \includegraphics[width=0.97\textwidth]{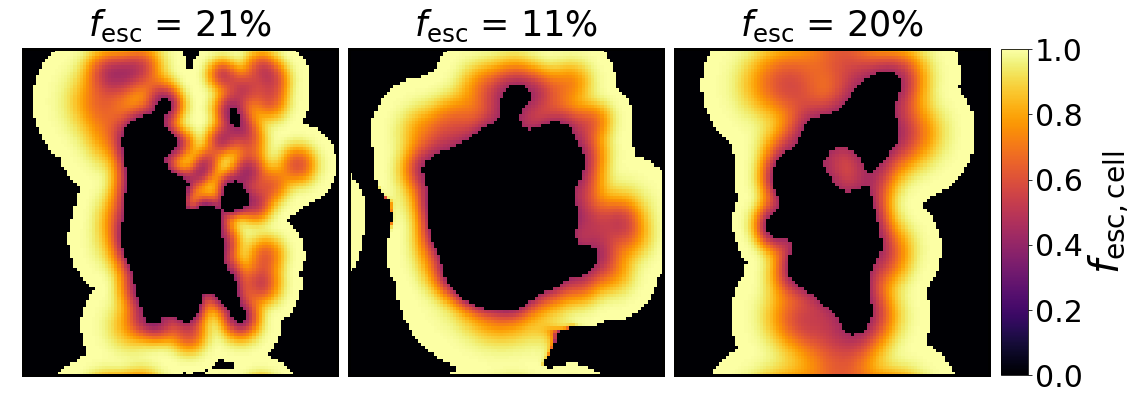}
    \includegraphics[width=0.94\textwidth]{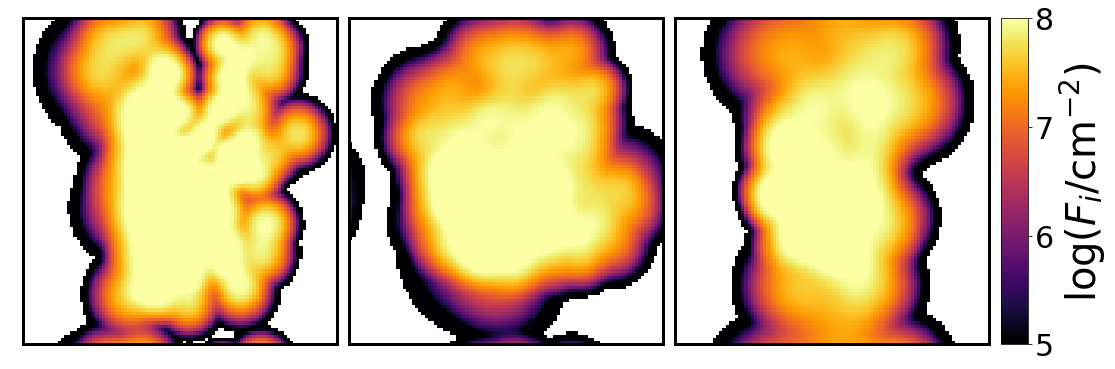}
    \includegraphics[width=0.98\textwidth]{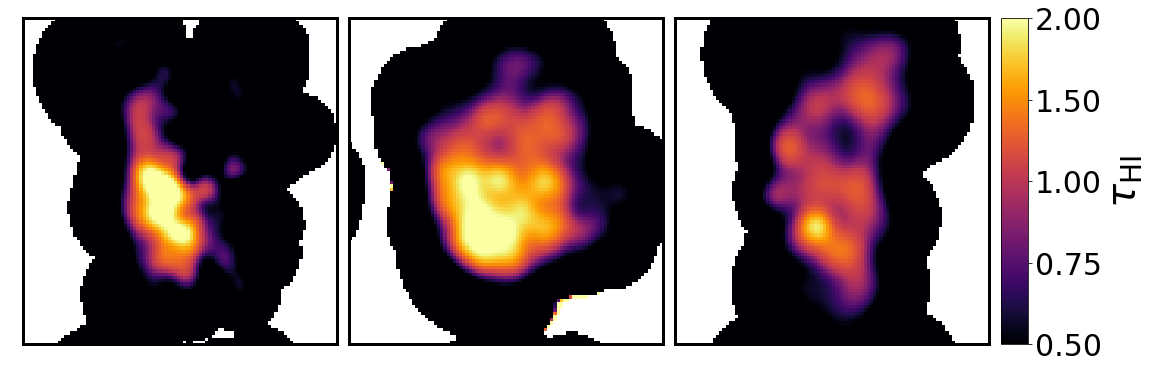}
    % \hspace{0.3cm}
    \end{subfigure}

\begin{subfigure}[t]{0.45\textwidth}
\caption*{LyC dark mode}
\includegraphics[width=0.97\textwidth]{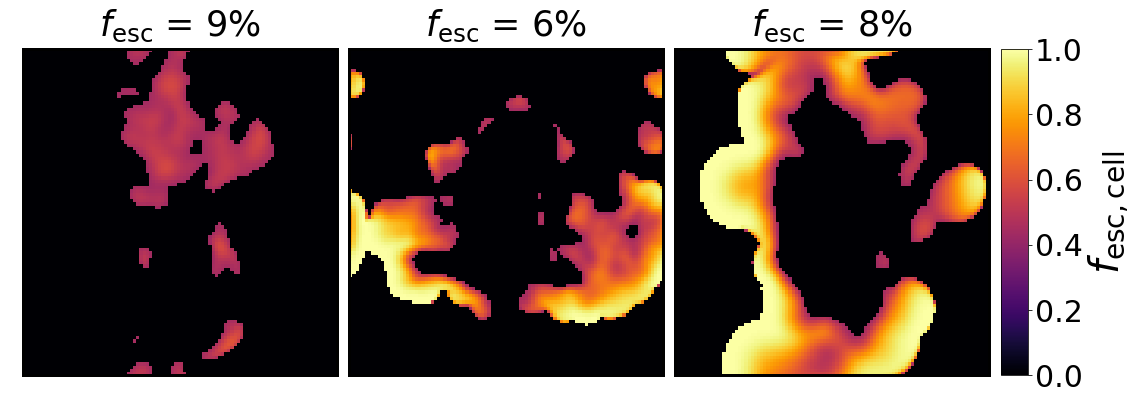}
  \includegraphics[width=0.96\textwidth]{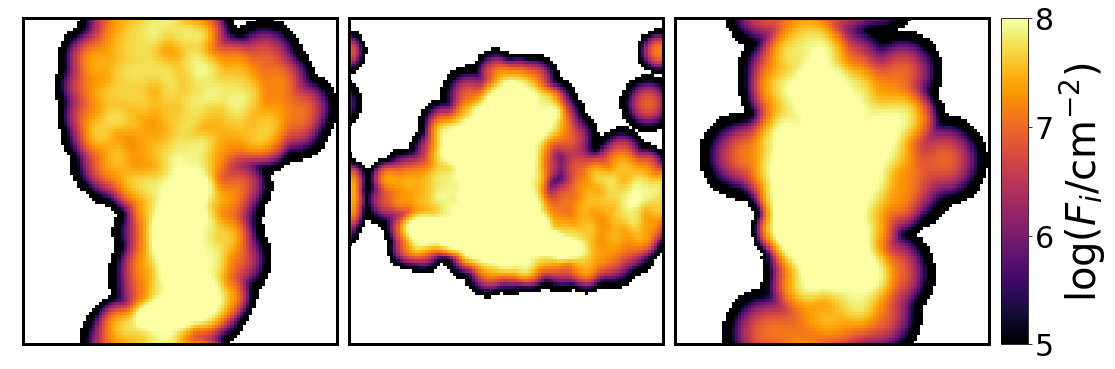} \\
  \hspace{0.2cm}
    \includegraphics[width=0.98\textwidth]{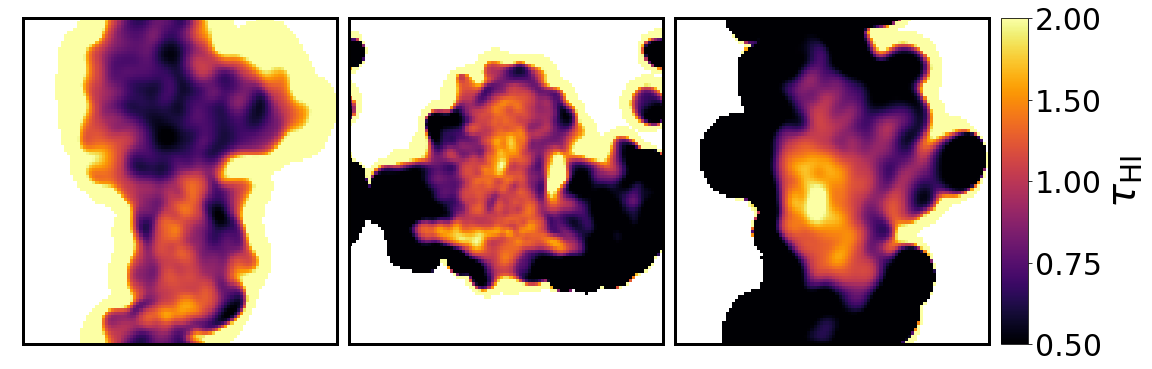}
    \end{subfigure}
    \caption{Projected maps of the escape fraction, ionizing flux, $F_i$, and the gas optical depth $\tau_\mathrm{HI}$, distribution from individual grid cells (box size $4$ stellar half mass radii) for three example galaxies from the \textit{loc}-mode (top left panels), \textit{ext}-mode (top right), and LyC dark mode (bottom). %\AF{I owuld remove the top text and instaed identify the three panels as loc-mode, ext-mode and Lyc-dark (?) mode - i.e. where the Lyc does not escape}{\ik Like this or should I also remove the 'Sample galaxies from' part.}
    %{\ik I think that is because in section 3.1 I used the terminology wrong. I changed it now and also adjusted the caption of fig. 2 a little bit. Conceptually the difference is simply that fesc is the escape fraction of one galaxy and <fesc> is the average escape fraction of all galaxies located in a bin of a 1 or 2D histogram. Changed the sample of galaxies selected from the central region.}
    }
    \label{fig:ns sample}
\end{figure*}

In fig.~\ref{fig:ns sample} we showcase a selection of three galaxies from each of the three regions marked in fig.~\ref{fig:Escape_columns_height}, by looking at the distribution of their escape fractions, their ionizing fluxes, $F_{\rm i}$, and the gas optical depth $\tau_\mathrm{HI}$, where the latter is approximately equivalent to the optical depth $\tau_\mathrm{tot}$, as in these cases $N_\mathrm{d}$ is negligible\footnote{We do not show $N_\mathrm{d}$ here as the map would essentially be fully dark.}.

Galaxies belonging to the \textit{loc}-mode exhibit relatively small localized escape regions, often situated in proximity to the center of the galaxy. Thus the earlier adopted name for said mode. In contrast, galaxies belonging to region 2 display large escape fractions over an extended area at the outskirts of the galaxies, with no LyC escaping from the inner regions. Finally, galaxies from the LyC dark mode display characteristics from both escape modes, but generally exhibit lower average escape fractions.

The \textit{loc}-mode reveals concentrated regions of ionizing flux, primarily found in the central areas of galaxies where the star formation process takes place. This results in the distinct compact escape channels. The abundance of ionizing photons in these regions leads to a low $\tau_\mathrm{HI}$ within the central areas of the galaxy. 
Consequently, the correlation between escape fraction and scale heights becomes apparent: as the density is inversely proportional to the scale height ($n \propto H^{-1}$), and the recombination rate increases $\propto n^2$, it follows that the escape fraction increases with increasing scale heights and hence less dense gas.

In contrast, galaxies in the \textit{ext}-mode exhibit a considerably larger area that emits ionizing flux. However, when analysing $\tau_\mathrm{HI}$, a noteworthy pattern emerges: the central region displays an inverse correlation between $F_\mathrm{i}$ and LyC escape, akin to what is observed in the \textit{loc}-mode.
Specifically, the central region exhibits the highest $\tau_\mathrm{HI}$ ratio, while the outskirts display minimal absorption. Consequently, in order for escape fractions to be substantial, it becomes crucial for a significant portion of star formation to occur in the outer regions of the galaxy. In this mode, the correlation with low scale height stems from star formation being driven by higher densities in the outer regions of the galaxy. Despite the recombination rate being larger in these higher density regions, it is compensated by the elevated production rate of ionizing photons driven by star formation. The interplay between star formation and gas absorption yields the distinct escape rings observed in galaxies sampled from the \textit{ext}-mode. These escape rings are a manifestation of the relative importance of star formation and gas absorption in shaping the escape characteristics of these galaxies.

Finally, galaxies from the boundary region exhibit properties that are observed in both the aforementioned escape modes. However, this regime presents limitations for escape from the center due to low scale height, as well as a scarcity of star formation in the outer regions of the galaxy. These factors collectively restrict escape through both channels, as illustrated in the lower panels of fig.~\ref{fig:ns sample}. The areas emitting ionizing flux in galaxies from the boundary region are less extended compared to the example galaxies from the \textit{ext}-mode. Additionally, there is a weaker correlation between low $\tau_\mathrm{HI}$ and high star formation regions. 
In summary, galaxies in the LyC dark mode show star formation characteristic of both escape modes, yet they lack the necessary conditions to facilitate LyC escape through either of these modes.

To verify the universality of the escape modes observed in our sample, we analysed the LyC escape characteristics within each galaxy. To quantify the degree of localization of the LyC escape in a given galaxy, we employed the Gini coefficient, which we calculate as
\begin{equation}
    G = \frac{2\sum_\mathrm{cell} i_\mathrm{cell} y_\mathrm{cell}}{N\sum_\mathrm{cell} y_\mathrm{cell}}-\frac{N+1}{N}, 
\end{equation}
with $N$ being the total number of grid cells in a given galaxy map and $y_\mathrm{cell} := f_\mathrm{esc, cell}F_\mathrm{ion, cell}$ the ionizing flux weighted escape fraction of each cell and $i_\mathrm{cell}$ being the index of a grid cells ordered by their $y_\mathrm{cell}$ values. A Gini value of 1 signifies that all ionizing radiation within a galaxy is concentrated in just one cell, resulting in a highly localized escape. On the other hand, a Gini value of 0 indicates that the ionizing photons leak equally from all cells, indicating a more evenly distributed escape pattern.

For galaxies exhibiting $f_\mathrm{esc}>0.1$ and a scale height $H<10^{20.8}\mathrm{cm}$ (corresponding to the \textit{ext}-mode), the average Gini coefficient is $\bar{G}_\mathrm{ext} = 0.7853 \pm 0.0010$. In contrast, galaxies with $f_\mathrm{esc}>0.1$ and a scale height $H>10^{21.3}\mathrm{cm}$ (corresponding to the \textit{loc}-mode) exhibit a higher average Gini coefficient of $\bar{G}_\mathrm{loc} = 0.9324 \pm 0.0002$. These findings clearly indicate a notable disparity in the LyC escape geometry between the two modes, and confirm our more qualitative previous discussion.

\begin{figure}
    \centering
    \includegraphics[width=0.47\textwidth]{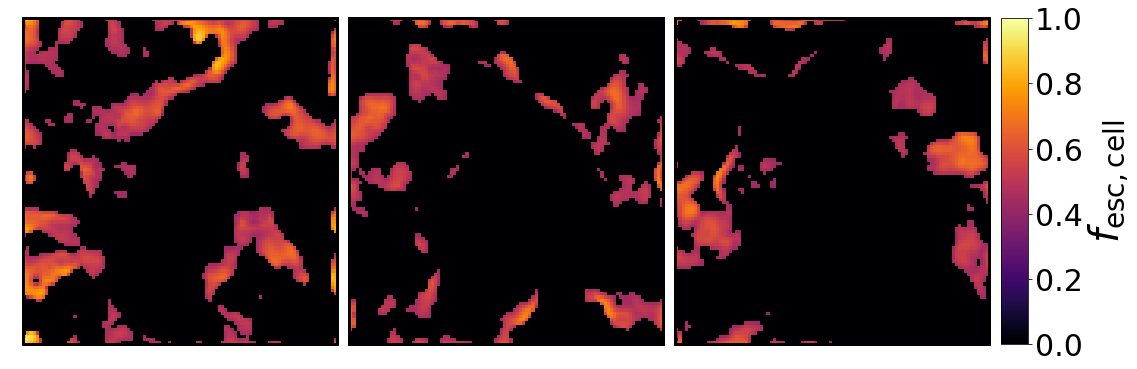}
    \includegraphics[width=0.49\textwidth]{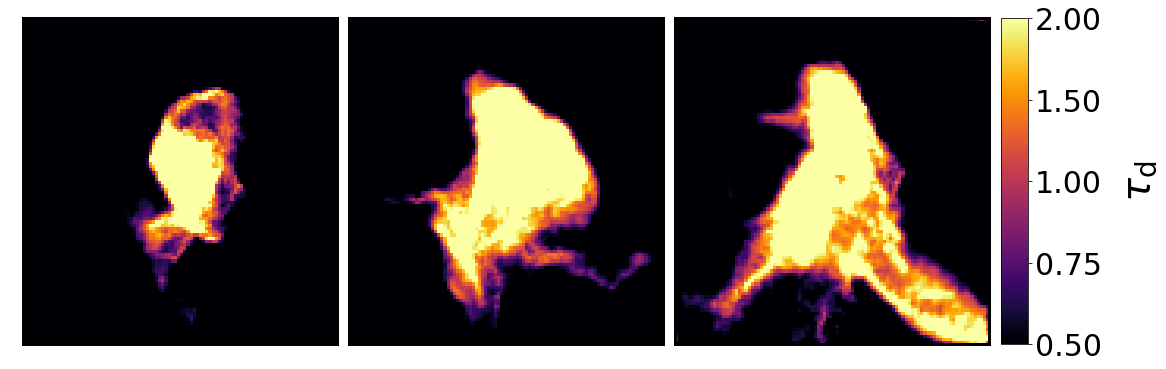}
    \includegraphics[width=0.48\textwidth]{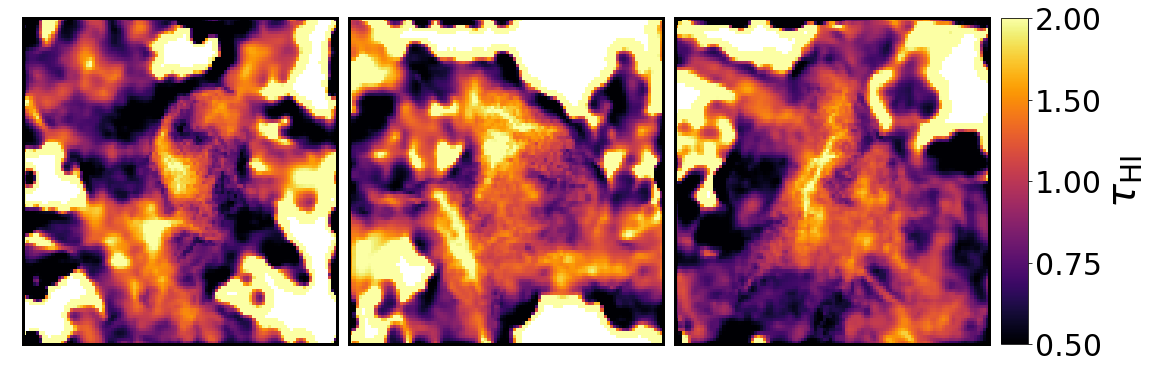}
    \vspace{0.5cm}

    \includegraphics[width=0.47\textwidth]{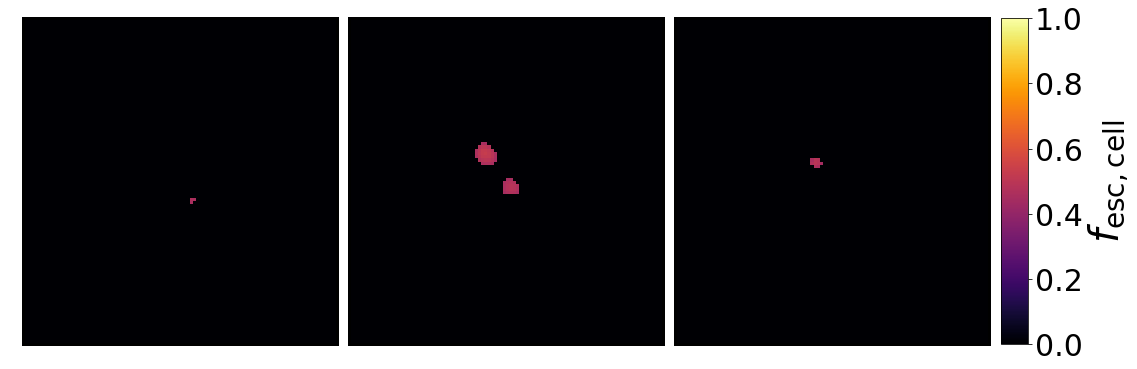}
    \includegraphics[width=0.49\textwidth]{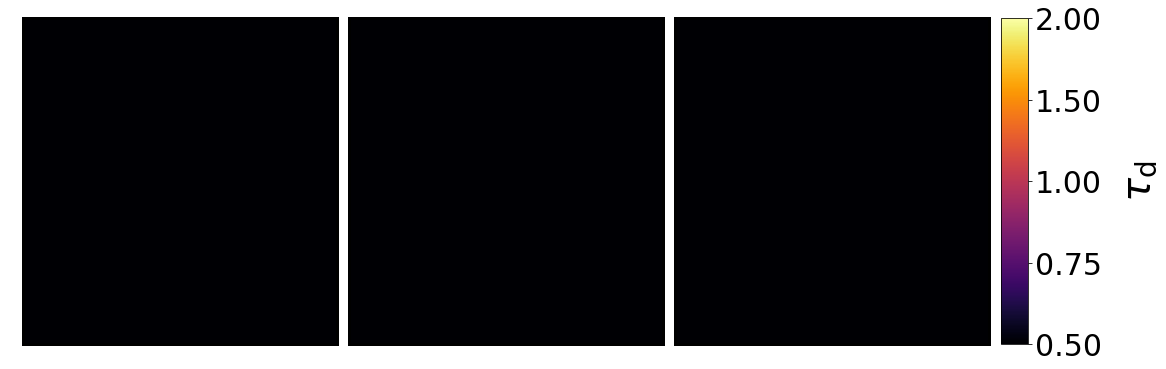}
    \includegraphics[width=0.48\textwidth]{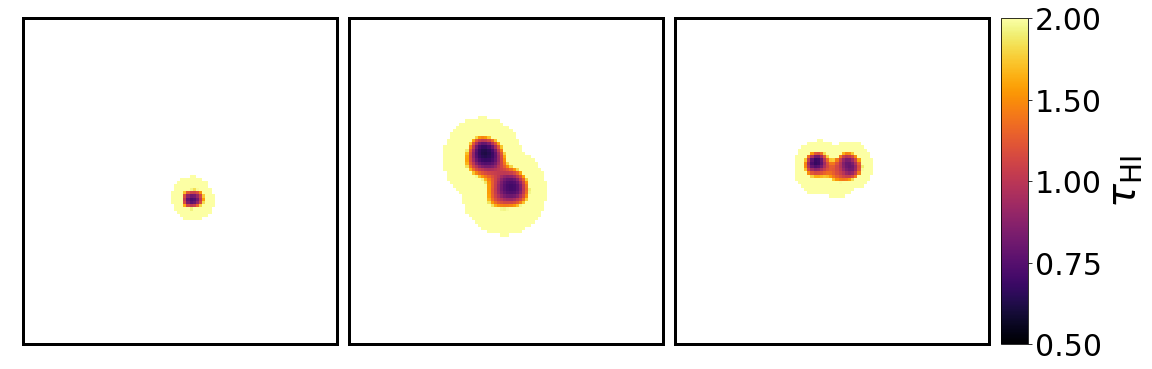}
    
    \caption{Galaxies with scale heights $H>10^{21.3}\mathrm{cm}$ selected for their high and low SFR, i.e. $\log(\mathrm{SFR}/\mathrm{M}_\odot \mathrm{yr}^{-1})=0.14, 0.20$ and $0.40$ (top) and $\log(\mathrm{SFR}/\mathrm{M}_\odot \mathrm{yr}^{-1})=-4.34, -4.05$ and $-4.08$ (bottom figure). From top to bottom, the panels in each figure refer to the cell escape fraction, $f_{\rm esc,cell}$, optical depth of dust, $\tau_\mathrm{d}$, and of gas, $\tau_\mathrm{HI}$. 
}
    \label{fig:properties bounds}
\end{figure}

In fig.~\ref{fig:Escape_columns_height} we noticed that in the \textit{loc}-mode for $\log (\mathrm{SFR})<-2.5$ the escape fraction is positively correlated to the SFR. However, for $\log( \mathrm{SFR})>-2$ the trend is reversed. To understand this, in fig.~\ref{fig:properties bounds} we examine in more detail galaxies selected on the basis of their SFR. In particular, we select three sample galaxies with high (i.e. $\log(\mathrm{SFR})>0$), and low (i.e. $\log(\mathrm{SFR})<-4$) SFR respectively, to explore the physical properties dominating the escape fraction in these regimes. 
In galaxies with high SFR, the determining factor for the escape fraction in the central region is the dust optical depth $\tau_\mathrm{d}$, which remains unaffected by the ionizing flux. 
However, $\tau_\mathrm{d}$ exhibits a correlation with the metallicity. Indeed, a higher metallicity content promotes gas cooling, resulting in an increase in SFR and subsequently leading to a decrease in the average escape fraction $\langle f_\mathrm{esc} \rangle$ with SFR. It should be noted, however, that as no dust destruction is present in our model, dust absorption should be regarded as an upper limit.
It is worth mentioning that galaxies at the upper limit in the \textit{loc}-mode showcase substantial LyC escape in the outer regions. Nevertheless, despite the extensive nature of these escape regions, only a small portion of the galaxies' ionizing photons are generated there, ultimately resulting in relatively low overall escape fractions.
Conversely, in the low SFR sample (bottom), we observe the presence of only a few small escape channels. In this scenario, the impact of dust absorption is negligible, while the LyC absorption is primarily influenced by the gas optical depth $\tau_\mathrm{HI}$. The escape fraction is thus strongly dependent on the number of ionizing photons generated, as $\tau_\mathrm{HI}$ is related to the ionizing flux (see eq.~\ref{eq:ns}), and consequently the SFR.

\begin{figure*}
    \centering
    \Large{\textbf{\textit{ext}-mode} \hspace{7cm} \textbf{\textit{loc}-mode}}
    \\
    \includegraphics[width=0.48\textwidth]{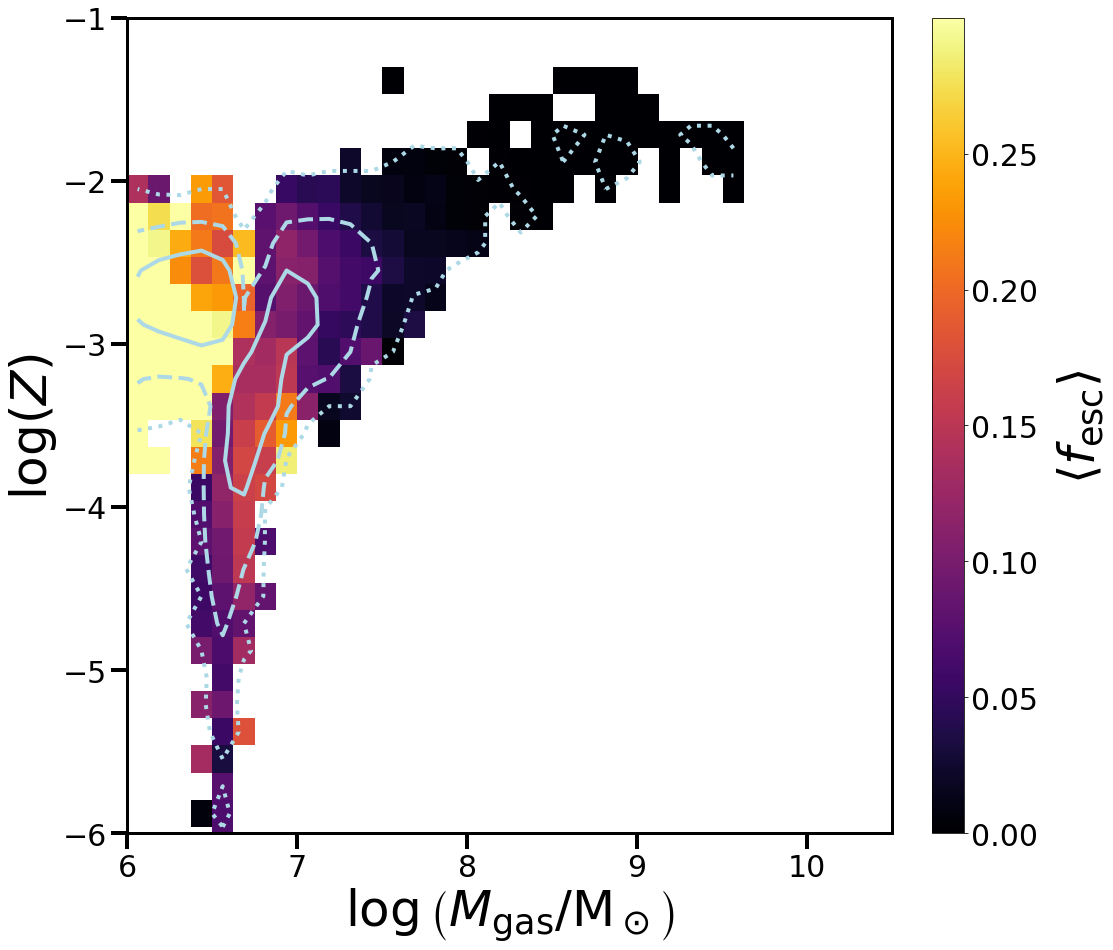}
    \includegraphics[width=0.48\textwidth]
    {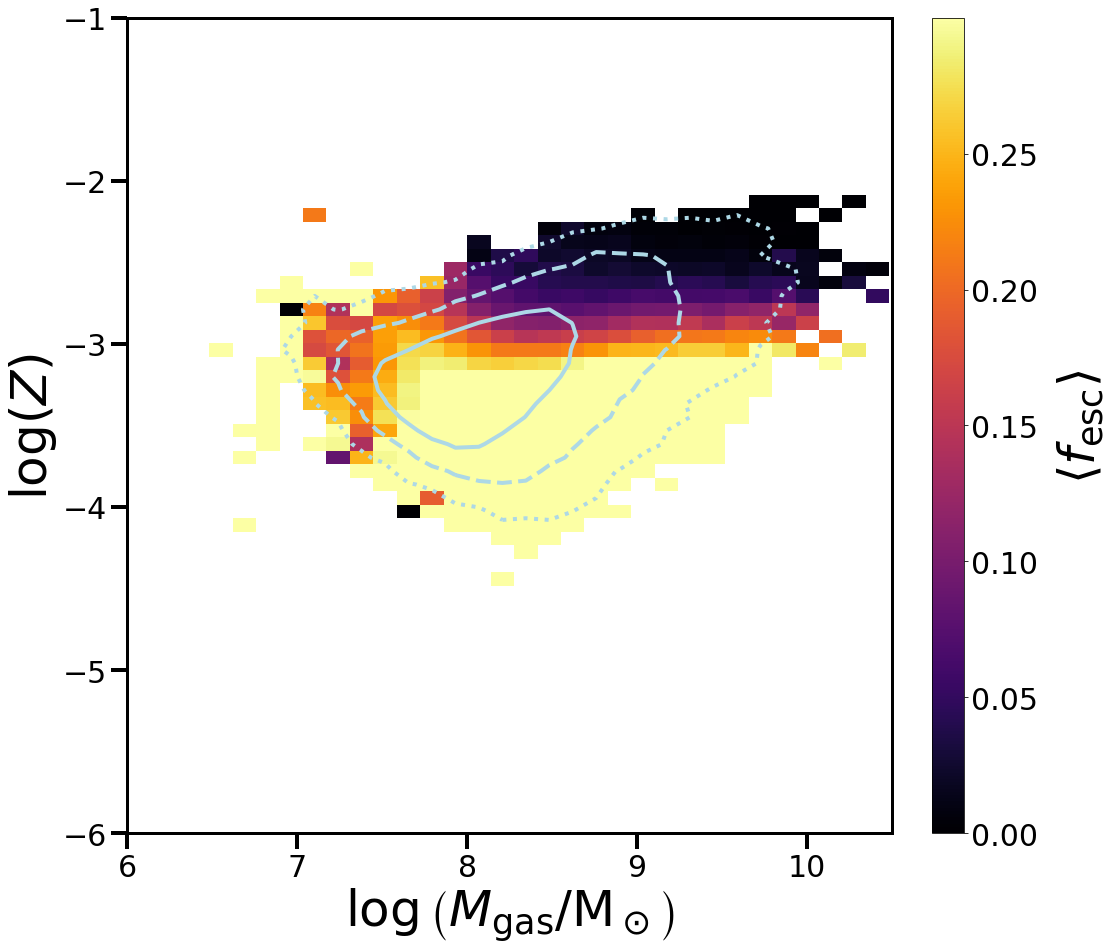}
    \includegraphics[width=0.48\textwidth]{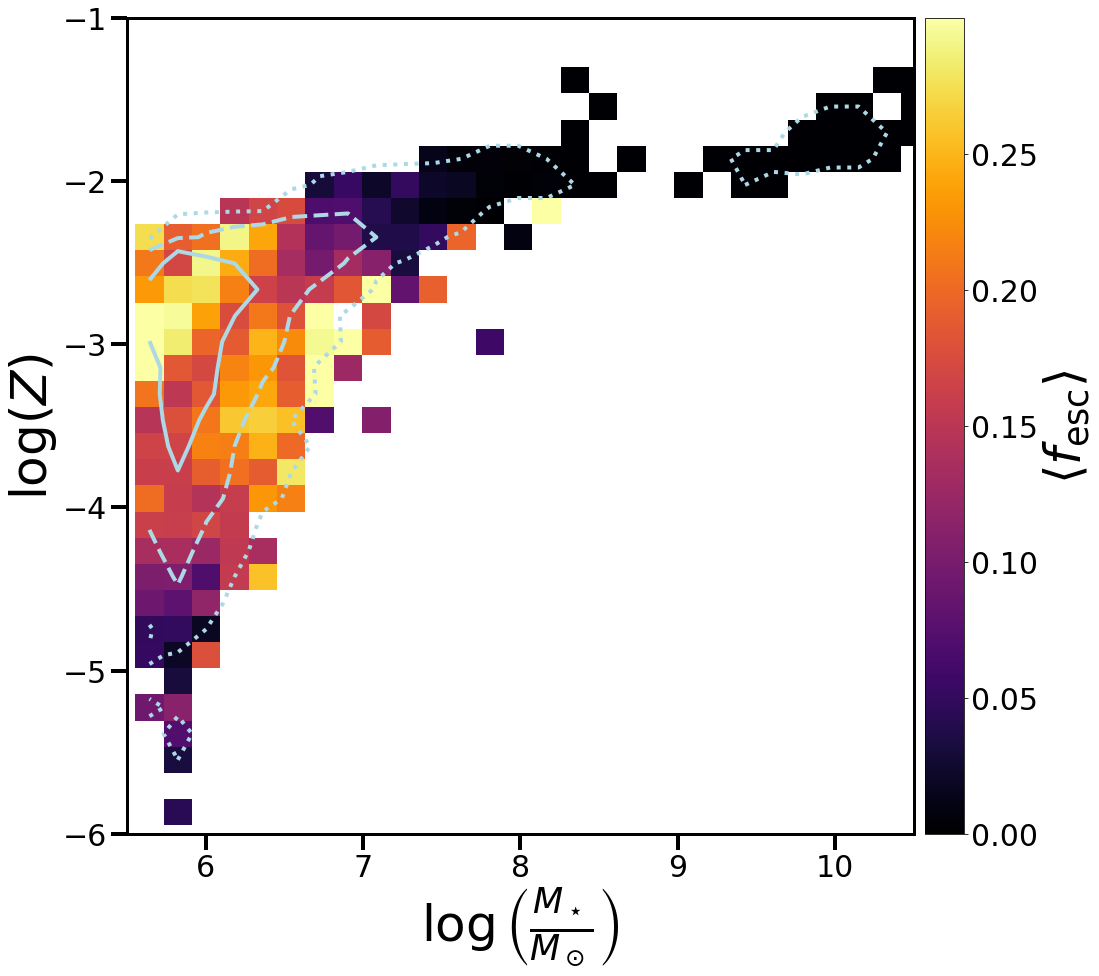}
    \includegraphics[width=0.48\textwidth]{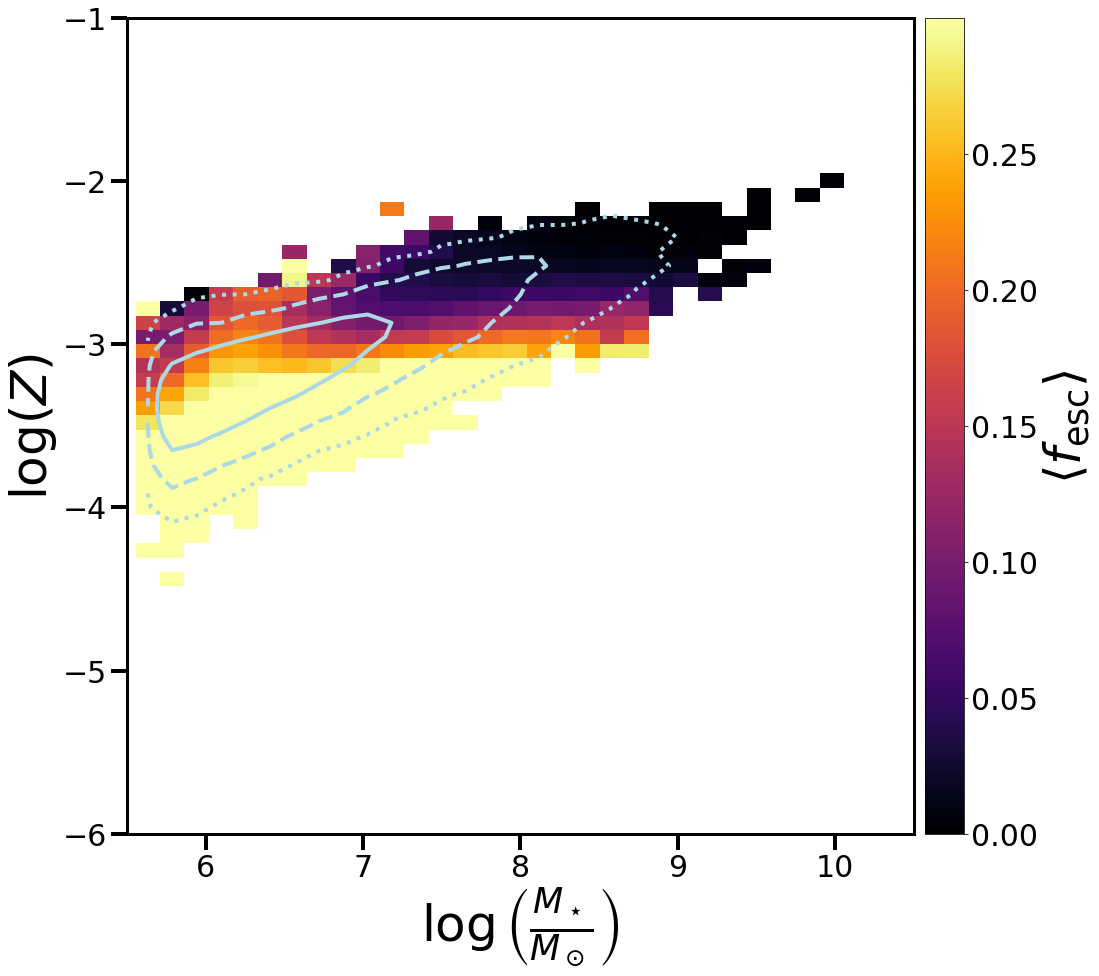}
    \includegraphics[width=0.48\textwidth]
    {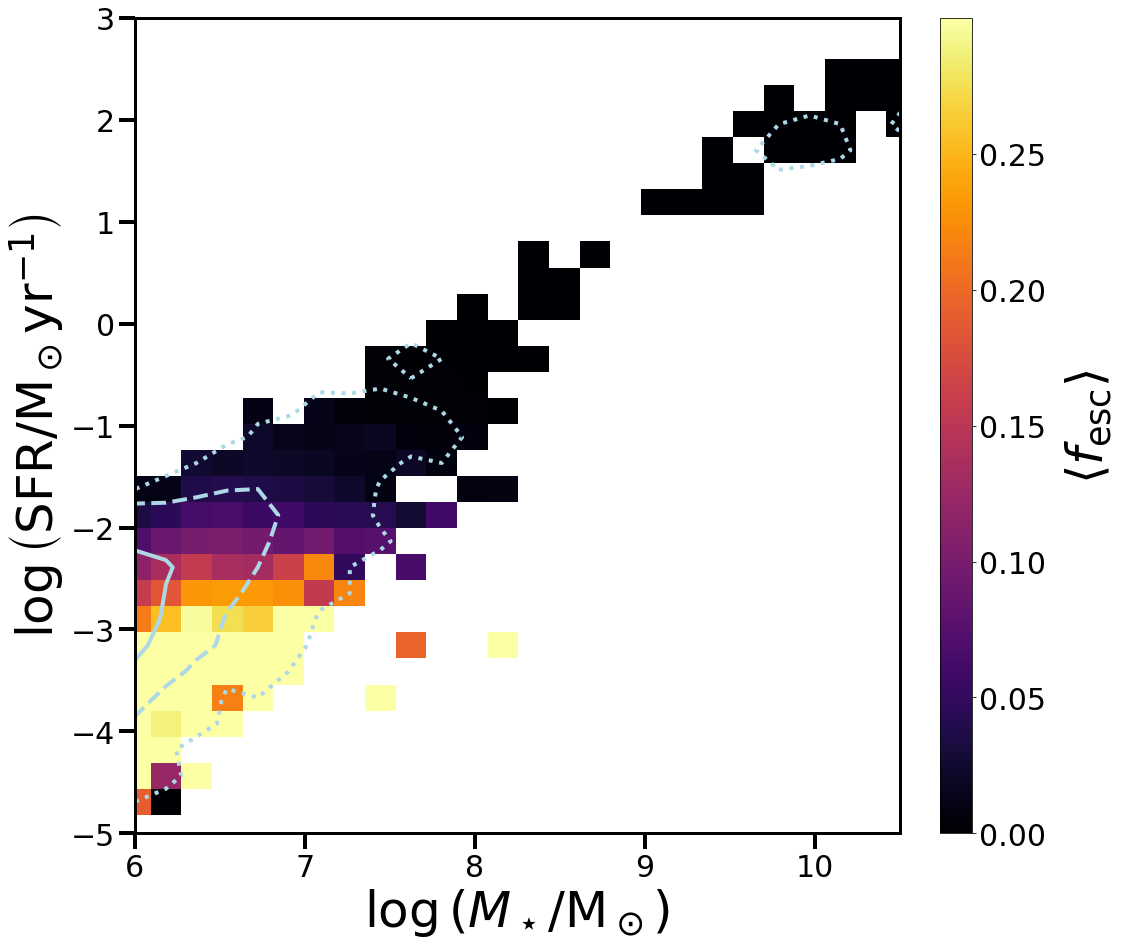}
    \includegraphics[width=0.48\textwidth]{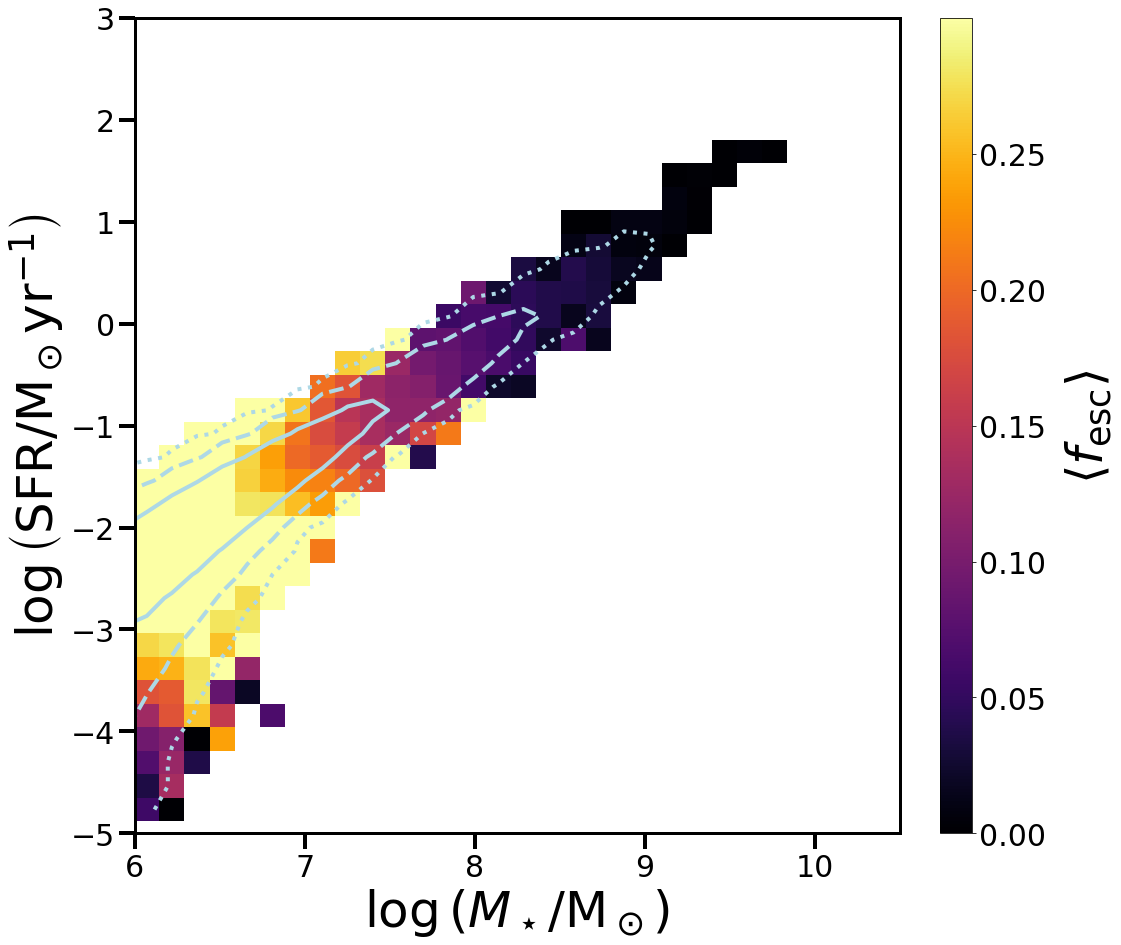}
    \caption{Average escape fraction as a function of: metallicity and galactic gas mass (top panels), metallicity and galactic stellar mass (center), and SFR and stellar mass (bottom). The left and right columns refer to galaxies with $\log (H/\mathrm{cm}) < 20.8$ and $\log (H/\mathrm{cm}) > 21.3$, which include the \textit{ext}- and \textit{loc}-modes respectively. The contours denote the 1  (solid), 2 (dashed) and 3 (dotted) sigma distribution of galaxy counts.
    }
    \label{fig:properties escape modes}
\end{figure*}

Finally, in fig.~\ref{fig:properties escape modes} we explore the correlation of the two escape modes with other galactic properties. For this purpose we separate the sample of galaxies into two groups: one with  $\log (H/\mathrm{cm}) > 21.3$ containing the \textit{loc}-mode, and one with $\log (H/\mathrm{cm}) < 20.8$ containing the \textit{ext}-mode. 
We then examine the correlation of these groups separately in relation to various galactic properties. 

From the top panels we see that there is a significant difference between the two groups with respect to dependencies on galactic gas mass and metallicity. 

For small scale heights (\textit{ext}-mode), $\langle f_\mathrm{esc} \rangle$ does not significantly depend on the metallicity, but is highly sensitive to the gas mass. 
All galaxies with significant escape fractions have $10^{-4}<Z<10^{-2}$ and $M_\mathrm{gas}\lesssim 10^7$~M$_\odot$. We can thus conclude that the \textit{ext}-mode is characterized by low gas mass, in addition to low scale heights and SFR, indicating that the bulk of the galaxies in this mode are limited in their LyC escape by gas absorption $\tau_\mathrm{HI}$.

Finally, within the sample of galaxies with a gas mass of approximately $10^{6.5}$M$_\odot$, we observe an evolution of $\langle f_\mathrm{esc} \rangle$ in relation to metallicity. For $Z<10^{-3.5}$, $\langle f_\mathrm{esc} \rangle$ increases with increasing metallicity, as in this regime dust absorption is expected to have minimal impact, while the enhancement of metal line cooling facilitates star formation, thereby contributing to a higher escape fraction. However, for $Z>10^{-3}$, this trend reverses due to the growing significance of dust absorption, which outweighs the additional star formation resulting from more efficient cooling.

Within the group encompassing galaxies with large scale heights \textit{(\textit{loc}-mode)}, a strong negative correlation between the average escape fraction $\langle f_\mathrm{esc} \rangle$ and metallicity becomes evident. Consequently, galaxies with $Z>10^{-3}$ exhibit negligible $\langle f_\mathrm{esc} \rangle$. This result highlights that the escape of LyC photons in the \textit{loc}-mode is predominantly limited by dust absorption, as exemplified in the galaxy sample depicted in fig.~\ref{fig:properties bounds}. Notably, there appears to be no upper limit to $f_\mathrm{esc}$ based on gas mass, indicating that the additional star formation facilitated by the gas compensates for the heightened gas absorption.
Conversely, in the case of galaxies with $M_\mathrm{gas}\lesssim 10^{7.5}$ M$_\odot$, the escape fraction experiences a decline, which reflects the limited star formation observed in these gas-deficient galaxies.

When investigating the average escape fraction as a function of metallicity and stellar mass instead of gas mass, notable changes occur in the distributions. The primary reason for this is the correlation between stellar mass and metallicity, which consequently correlates with the dust optical depth ($\tau_\mathrm{d}$).
For galaxies with small scale heights (\textit{ext}-mode), the bins displaying high $\langle f_\mathrm{esc} \rangle$ values span a wider region in the $M_\star-Z$ space. This indicates that dust absorption plays a secondary role in the LyC escape for these galaxies. Nonetheless, the average escape fractions are generally lower compared to the plot above, as galaxies within a specific stellar mass bin exhibit varying gas fractions and thus have a wider spread in $\tau_\mathrm{HI}$.
For galaxies with large scale heights, we observe that those with $M_\star \gtrsim 10^8\mathrm{M}_\odot$ have negligible escape fractions. This results from the high metallicities ($Z>10^{-3}$), leading to strong dust absorption and consequently smaller escape fractions.

When examining the distribution of galaxies as a function of stellar mass and SFR, we see that both groups have a similar behaviour. However, the group with low scale heights (\textit{ext}-mode) has a slightly higher scatter in SFR at a fixed $M_\star$. This is likely caused by both old, mostly quenched galaxies with little feedback, as well as efficiently cooling high SFR galaxies with small scale height. 
The sequence corresponding to the small scale heights extends to approximately one order of magnitude higher SFR and $M_\star$. 
The reason is that large galaxies are usually found at lower redshifts and are usually more enriched by metals. This facilitates their cooling resulting in lower scale heights.

\subsection{Redshift evolution} \label{sec:semi_redshift}
\begin{figure}
    \centering
    \includegraphics[width=0.5\textwidth]{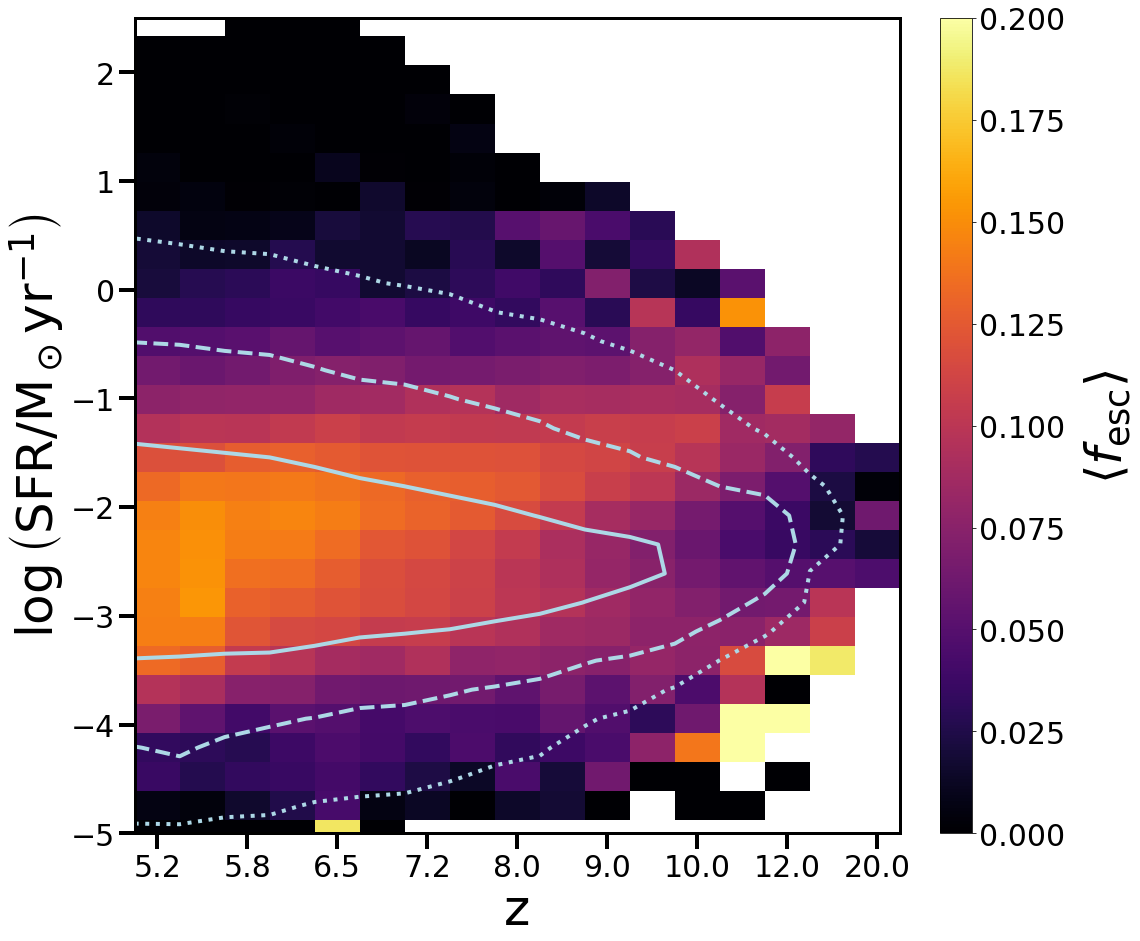} \\
    \includegraphics[width=0.5\textwidth]{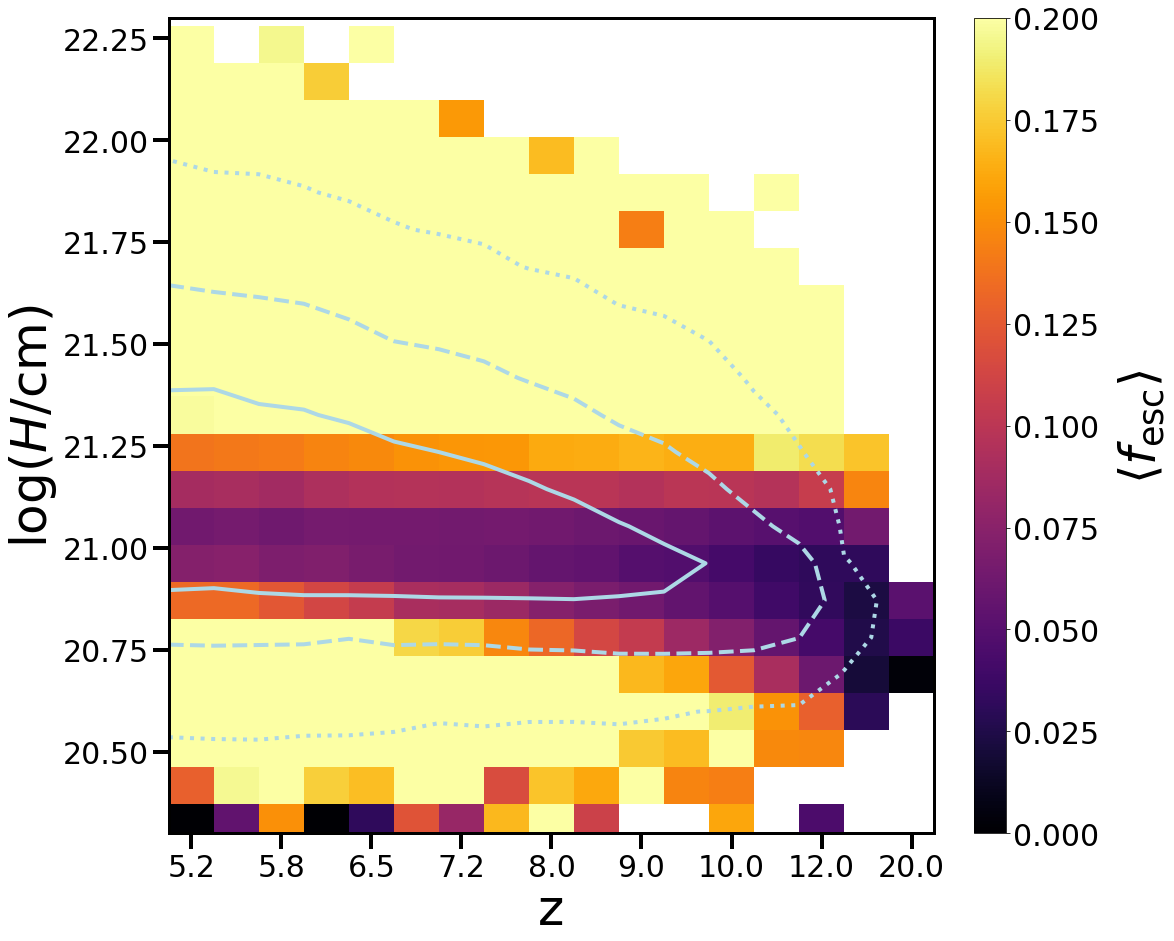}
    \caption{Redshift evolution of the average escape fraction as a function of SFR (upper panel) and scale height (lower). The contours
denote the 1 (solid), 2 (dashed) and 3 (dotted) sigma distribution of galaxy
counts.}
    \label{fig:evolution with z}
\end{figure}

In fig.~\ref{fig:evolution with z} we examine how the dependence of $f_\mathrm{esc}$ on SFR and scale height discussed in the previous section evolves with redshift.
In the top panel we observe that the highest values of $\langle f_\mathrm{esc} \rangle$ are found at higher values of SFR with increasing redshift, resembling the dependence on $M_\star$ depicted in fig.~\ref{fig:general histograms}. To understand this trend we note that at higher redshifts galaxies are more metal poor. This means that escape through the localized mode is facilitated, while escape through the extended mode is either unaffected or even hindered when the metallicity is not sufficient to facilitate the extended star formation required. As the \textit{ext}-mode is only found in galaxies with $M_\star \lesssim 10^7\mathrm{M}_\odot$, the lack of this escape mode at higher redshifts results in $\langle f_\mathrm{esc} \rangle$ peaking at higher stellar masses.

This behaviour is further supported by the lower plot, where the $\langle f_\mathrm{esc} \rangle$ for galaxies with small scale heights, i.e. corresponding to the \textit{ext}-mode, decreases with increasing redshift within the range of $10^{20.5}$cm$<H<10^{21}$cm.
Conversely, for galaxies with large scale heights exhibiting localized LyC escape, $\langle f_\mathrm{esc} \rangle$ shows an opposite trend, particularly evident in the range of $10^{21.1}$cm$<H<10^{21.3}$cm.
As a result, as seen in fig.~\ref{fig:line plots}, the LyC escape fraction averaged over all galaxies decreases with increasing redshift due to less massive galaxies not leaking photons through the process of the \textit{ext}-mode. Even though leakage through the \textit{loc}-mode for massive galaxies is enhanced at high redshifts, their abundance in the early Universe is too low, to compensate for the decrease in the \textit{ext}-mode leakage.

While this is surprising given the strong observational evidence for a decrease in LyC escape with redshift, our results do not contradict these findings. First, the observed reduction in escape fraction with redshift is mostly seen in the \textit{ext}-mode, which is predominant in low mass galaxies. Instead, galaxies identified as LyC leakers have higher stellar masses, likely categorizing them into the \textit{loc}-mode, where we observe an increase in the escape fraction with redshift. Additionally, considering that a UV-background in TNG50 is turned on only at $z=6$, we can expect an  underestimation of quenching effects on low-mass galaxies, potentially leading to an overestimate of their LyC leakage. We plan to confirm this behaviour by applying the method also to other simulations.

\section{Fitting model} \label{sec:semi_fitting}
Using a more precise subgrid model of LyC escape can enhance the accuracy of future large-scale cosmological models of reionization since large-scale simulations cannot cover the full range of scales relevant to the radiation escape process. For this purpose, we develop a fitting function which allows to predict the escape fraction based on a few macroscopic properties of galaxies.

In order to model the escape fraction, we use a polynomial fitting method with three galactic properties, namely the stellar mass of the galaxy $M_\star$, its gas mass $M_\mathrm{gas}$, the metallicity $Z$, as well as the redshift $z$. We used a quadratic model and found that there is little change in the accuracy when increasing the number of parameters beyond seven. We also found that metallicity does not appear in the best fit model, which is given by
\begin{equation}
\begin{split}
    \tilde{f}_\mathrm{esc} = & 0.1197 x_\star^2 + 0.3737 x_\mathrm{gas}^2 - 0.4676 x_\star x_\mathrm{gas} \\ &- 0.0019 z x_\mathrm{gas}
        + 1.8589  x_\star - 2.5041  x_\mathrm{gas} + 3.5620,
\end{split}
\end{equation}
where the masses are expressed in logarithmic units, i.e. $x_{\alpha} = \log_{10} (M_\alpha/\mathrm{M}_\odot)$, and $\tilde{f}_\mathrm{esc}$ is the unconstrained estimate for the LyC escape fraction.
However, some of the values of the unconstrained estimate are below 0 or above 1, and hence outside the physically possible range. 
These results should then be set to the lower bound of 0 and upper bound of 1, respectively. Finally, since only a few galaxies in our sample have  $x_\star>8.5$, the estimates in this range are likely to be inaccurate. Indeed, the unconstrained estimate $\tilde{f}_\mathrm{esc}$ predicts high escape fraction values in the high stellar mass range and hence does not match our findings in fig.~\ref{fig:Escape_columns_height}. We therefore artificially set the $f_\mathrm{esc}$ value to 0 for galaxies with $x_\star>8.5$, obtaining:
\begin{equation}
f_\mathrm{esc} = \begin{cases}
0 \hspace{1cm} &\text{$\tilde{f}_\mathrm{esc}<0$ or $M_{\star,\mathrm{log}}<8.5$} \\
1 &\text{ $\tilde{f}_\mathrm{esc}>1$} \\
\tilde{f}_\mathrm{esc} &\text{otherwise}. \label{eq:fitting}
\end{cases}
\end{equation}

In order to determine the accuracy of the model in predicting individual escape fractions, we select a subsample of  galaxies not used for the fitting in order to avoid problems due to over-fitting. As a measure for this accuracy we use the average relative deviation
\begin{equation}
    r = \frac{1}{N_\mathrm{test}}\sum_{i=1}^{N_\mathrm{test}} \frac{\lvert f_{\mathrm{esc, pred},i}-f_\mathrm{{esc},i} \rvert}{f_{\mathrm{esc},i}},
\end{equation}
with $f_{\mathrm{esc, pred},i}$ and $f_{\mathrm{esc},i}$ being the predicted and modelled escape fraction of the $i$-th galaxy respectively, and $N_\mathrm{test}$ the number of test galaxies. We only used galaxies with $f_\mathrm{esc}>0.01$, as this measure is not useful for $f_\mathrm{esc}$ approaching 0. We find a value of $r \approx 1.2$, i.e. the average estimation error is of the order of a factor of 2, and as such the accuracy of predicting the escape fraction of a single halo is limited. However, for large scale studies where the statistical distribution of the escape fraction is more important, this model performs significantly better. Indeed, the average escape fraction obtained with the fitting formula is $\bar{f}_{\mathrm{esc, pred},i} = 0.121 \pm 0.086$, with the modelled escape fraction being $\bar{f}_{\mathrm{esc},i}=0.117 \pm 0.1334$.

\begin{figure}
    \centering
    \includegraphics[width=0.49\textwidth]{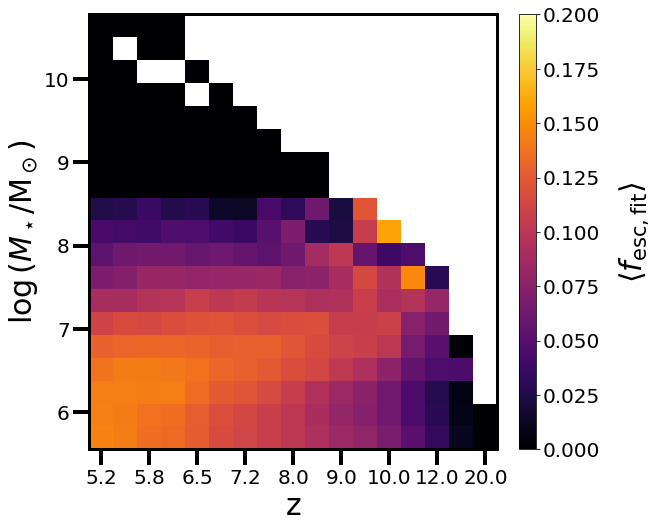}
    \caption{As the top panel of fig.~\ref{fig:general histograms} with the escape fractions obtained using the fitting formula in eq.~\ref{eq:fitting}.}
    \label{fig:fitting fesc mstar}
\end{figure}

In fig.~\ref{fig:fitting fesc mstar} we show how well the fitting formula is able to reproduce the behaviour of the escape fraction in relation to the stellar mass and redshift that we examined in fig.~\ref{fig:general histograms}. We see that the evolution of the escape fraction with redshift is successfully reproduced. However, the large gradients in $\langle f_\mathrm{esc} \rangle$ that are seen in fig.~\ref{fig:evolution with z} are smoothed out. 
The reason for this likely lies in the optimization process used to find the fitting formula, as the mean squared error was used for optimization, and thus large gradients in the fitting function were disfavored because they led to large errors for the outer mass ranges.

\begin{figure}
    \centering
    \includegraphics[width=0.49\textwidth]{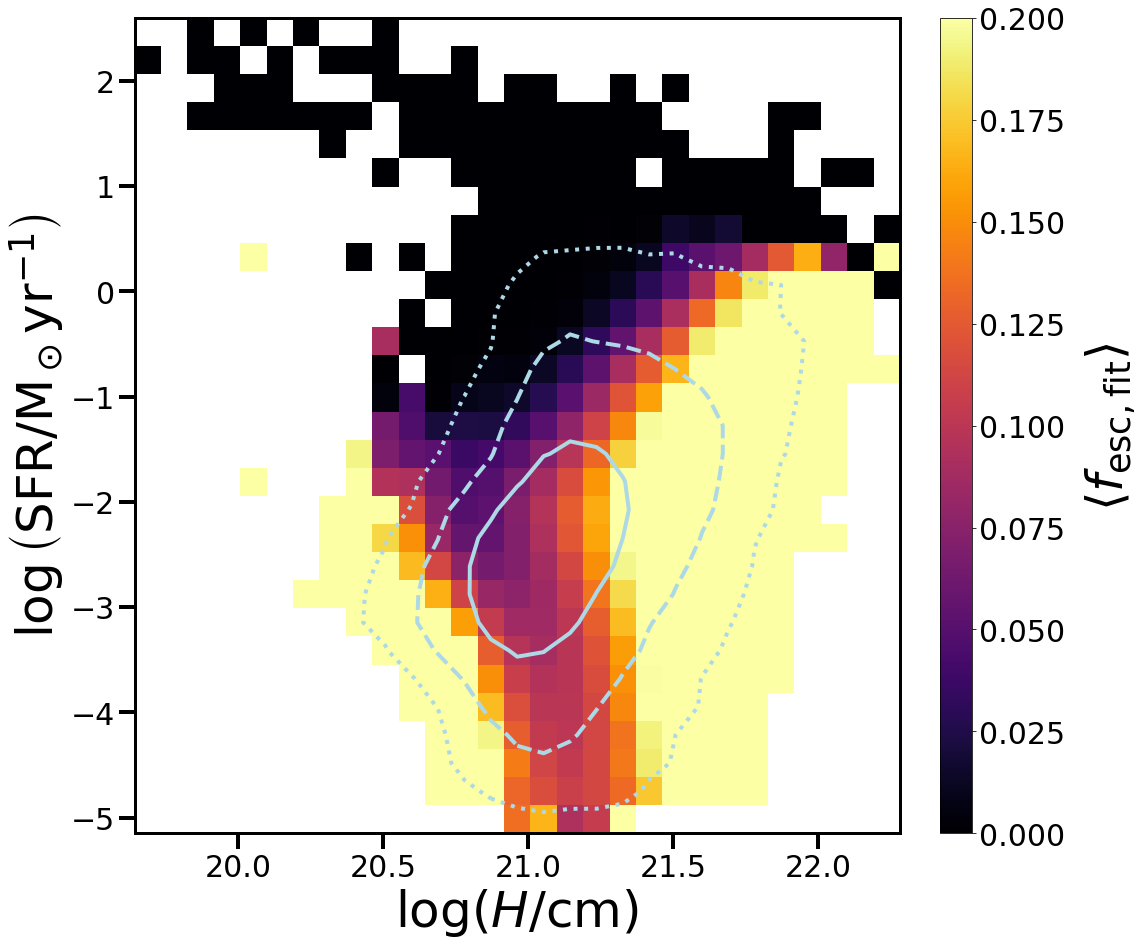}
    \caption{As the top panel of fig.~\ref{fig:Escape_columns_height} with the escape fractions obtained using the fitting formula in eq.~\ref{eq:fitting}.}
    \label{fig:fitting fesc height}
\end{figure}

Fig.~\ref{fig:fitting fesc height} shows that the fitting formula is able to successfully predict the bimodality in the escape fraction, as seen in fig.~\ref{fig:Escape_columns_height}. However the boundary between the two modes is less pronounced. This is likely caused by the smoothing effect of the optimization process of the fitting function discussed above.

\begin{figure}
    \centering
    \includegraphics[width=0.49\textwidth]{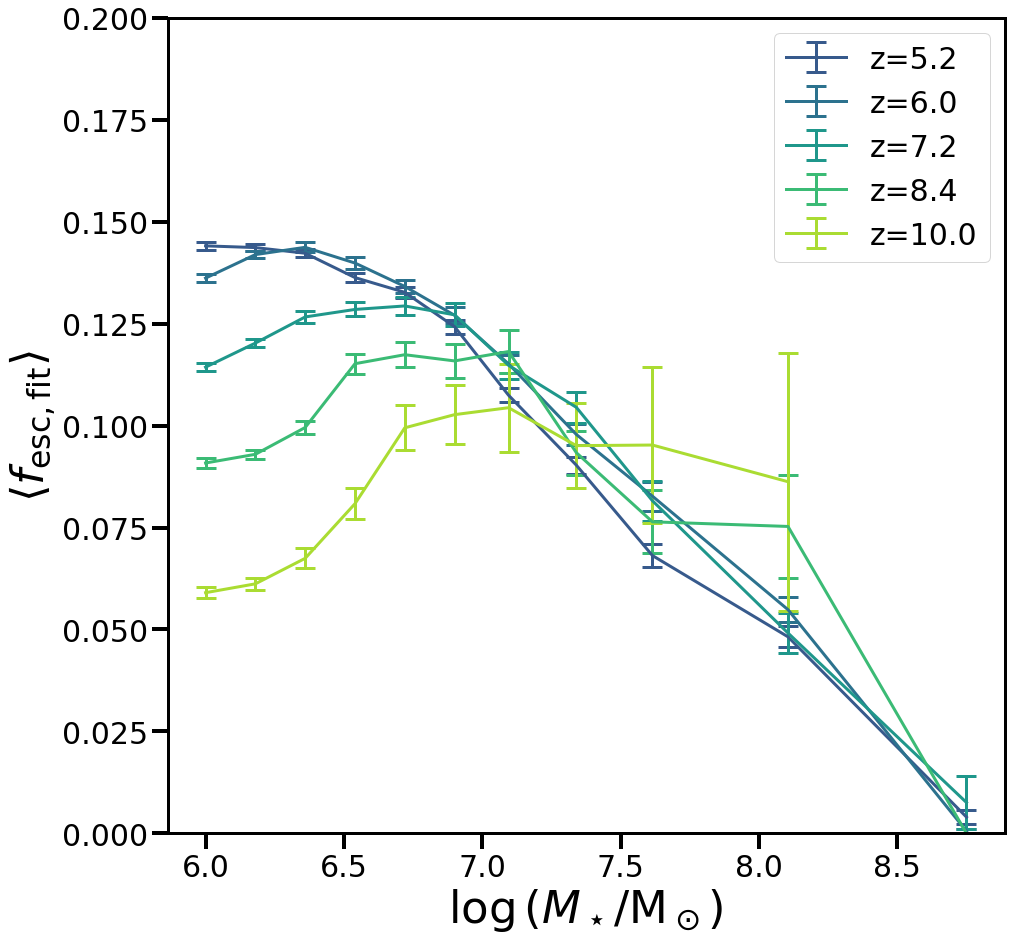}
    \includegraphics[width=0.49\textwidth]{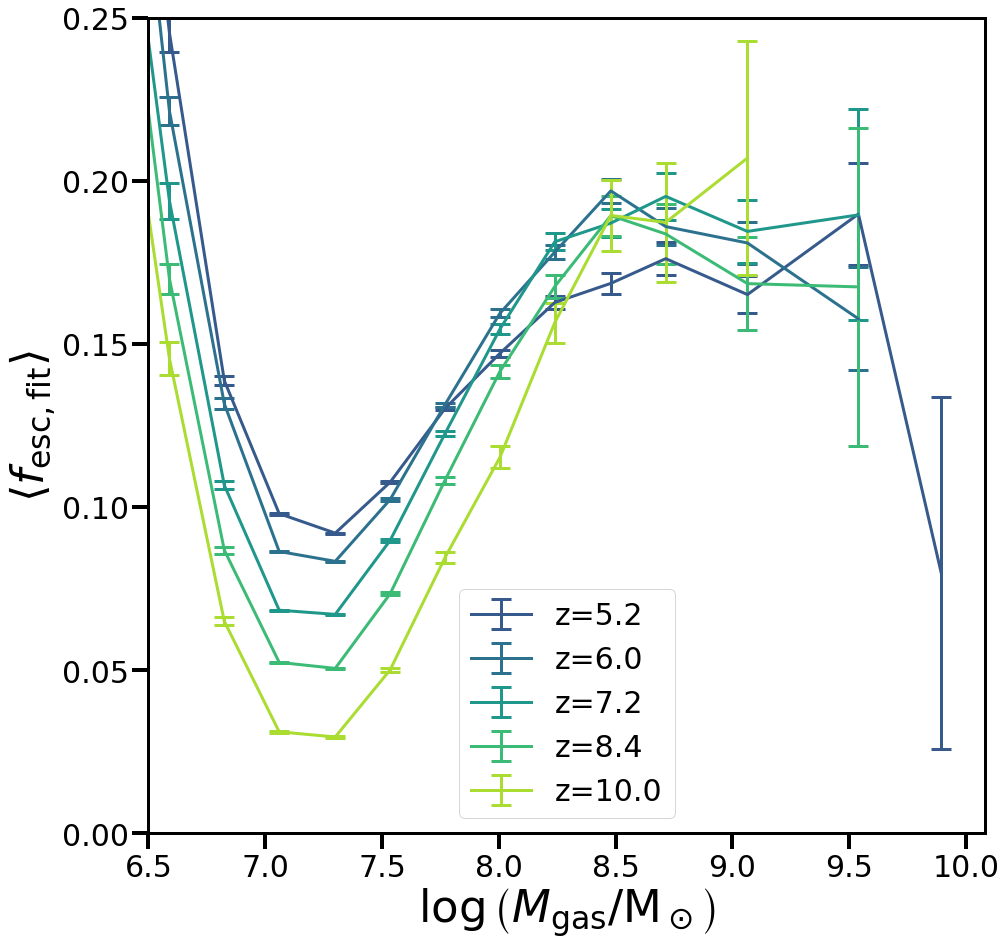}
    \caption{As fig.~\ref{fig:line plots} with the escape fractions obtained using the fitting formula in eq.~\ref{eq:fitting}.}
    \label{fig:fitting fesc lines}
\end{figure}

By comparing fig.~\ref{fig:fitting fesc lines} to fig.~\ref{fig:line plots}, we see that the fitting formula is able to reproduce all important trends, namely, the decrease in peak escape fraction with redshift and the approximate locations and values of the peaks. We also reproduce both the minima and maxima in the dependence of $f_\mathrm{esc}$ on $M_\mathrm{gas}$. 

\begin{figure}
    \centering
    \includegraphics[width=0.49\textwidth]{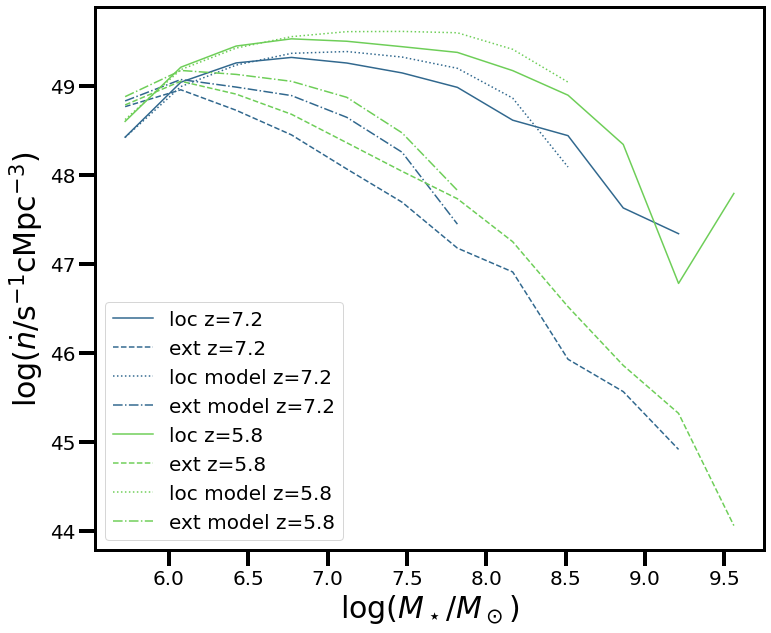}
    \caption{Comparison of the ionizing emissivity obtained using the semianalytic model as compared to the fitting formula at two sample redshifts}
    \label{fig:ndot_fit}
\end{figure}

Finally, in fig.~\ref{fig:ndot_fit} we compare the ionizing photon escape density obtained using the seminalytic model directly to the results from the semianalytic model. As one can see, the model is highly successful in matching the photon escape at lower stellar masses with higher discrepancies at higher masses. This can be explained given the high numerical abundance of these low-mass galaxies (see Fig.~\ref{fig:counts}) dominated the fitting process. In general, we can say that the fitting model overpredicts the total ionizing emissivity density by 10-20\%. This discrepancy is not noticeably dependent on redshift and is higher for the \textit{ext}-mode compared to the \textit{loc}-mode.

As mentioned earlier, it is important to emphasize that our modeling aims to capture the overall trends of LyC escape with galactic properties. Considering the inherent limitations in resolution and simplifications involved in estimating the LyC flux, it is crucial to scale the absolute value predicted by the fitting formula using a free parameter, which should be determined based on the specific ionizing photon budget required for reionization. We intend to investigate the large-scale implications of these results and to determine scaling parameters in subsequent work.

\section{Discussion and Conclusions} \label{sec:semi_conclusions}

To gain a better understanding of the correlation of LyC escape with galactic properties, we have applied the physically motivated model for the LyC escape fraction developed in F23 to $\approx$ 600,000 galaxies extracted from the TNG50 simulation \citep{nelson19a, pillepich19}  in the range $5.2<z<20$. It should be emphasized that, unlike in our previous study \citep{kostyuk23}, we have employed a very different source model. In this work, ionizing emissivity is sourced by star-forming gas rather than by earlier formed star particles. Hence, the ionizing emissivity traces the instantaneous star-formation rather than its recent history. Therefore, while the underlying simulation in both studies is the same, the difference in the source model strongly limits the correlation between the LyC escape fractions obtained using the two different approaches as discussed in detail in appendix~\ref{app:comparison_num}.

Given the large uncertainties in the subgrid modeling of LyC escape, attempting a quantitative comparison of our results to those of previous studies would be impractical. Therefore, we focus our discussion on qualitative results. Numerous previous numerical studies, such as the First Billion Year project \citep{paardekooper15}, FIRE-II \citep{ma20}, THESAN \citep{yeh22}, SPHINX \citep{rosdahl22} and TNG50 \citep{kostyuk23}, have consistently demonstrated an increase in the escape fraction with galactic stellar mass, with a maximum reached for galaxies with $M_\star \approx 10^7-10^8 \mathrm{M}_\odot$.
%Not all numerical results predict a steep decline in the escape fraction for higher stellar masses. However, SPHINX, CODA-II and, at some redshifts, THESAN do predict this trend as well.

The main difference from most previous investigations lies in the prediction of the average escape fraction increasing with decreasing redshift, which we find to be caused by a higher escape fraction from low-mass galaxies at lower redshifts. Furthermore, we find the presence of an additional extended escape mode for high redshift low-mass galaxies.
% to the emergence of the extended LyC escape mode in the galactic population with low stellar masses once sufficient metal enrichment has occurred. This result is typically found in observational investigations (e.g. \citealt{inoue06}), as well as theoretical ones (e.g. \citealt{rosdahl22, yeh22}).
A potential explanation for the absence of this mode in previous studies, including our own \cite{kostyuk23}, could be attributed to the predominant reliance on stellar particles as representations of stellar populations. Considering that star formation in the \textit{ext}-mode takes place across a vast region at a relatively gradual rate, relying on the sampling of relatively massive particles at fixed locations might not provide an accurate depiction of the production of ionizing photons in this mode (see appendix~\ref{app:comparison_num} for a comparison between the two methods). However, given the strong limitation of LyC escape due to dust in the \textit{loc}-mode we expect the escape fractions to decline again at even lower redshift. In the future, it will be critical to extend the study to lower redshifts, to confirm our findings by applying the method to different simulations, as well as investigate what they imply with respect to the reionization history.

% Nevertheless, the presence of additional quenching caused by a UV background and other simulation-specific modeling factors could influence the manifestation of this escape mode. Hence, further investigations are needed to explore the emergence of escape modes in alternative simulation setups. 

We note also that \cite{katz22b} identified two distinct modes of LyC escape in the SPHINX simulation, characterized as "bursty" and "remnant leakers," contingent on the galaxies' SFR. The former exhibit low SFR values and share a somewhat higher metallicity threshold with the \textit{ext}-mode in our study. However, they extend to significantly higher stellar masses than our \textit{ext}-mode. However, due to the different methodologies employed in these studies, a more quantitative analysis is necessary before a definitive comparison between the two findings can be established.

Our main results can be summarized as follows.
\begin{itemize}
    \item LyC escape is maximum for low-mass galaxies with $M_\star = (10^6-10^7) \mathrm{M}_\odot$ and then decreases, such that galaxies with $M_\star>10^9\mathrm{M}_\odot$ have negligible $f_\mathrm{esc}$ values.
      \item High redshift galaxies exhibit two main modes of LyC leakage:
(a) the {\it extended escape mode}  is characterized by extensive star formation regions, where LyC photons escape from the outer parts of galaxies (Gini score $0.79$) with low gas column density. Galaxies in this mode usually have a higher metal content ($10^{-3.5}<Z<10^{-2}$), which facilitates gas cooling and its collapse onto a thin plane with low scale heights ($H<8 \times 10^{20}$cm) where stars can form over a wide area. Here, dust absorption plays a subdominant role in limiting LyC escape, with respect to gas absorption. These galaxies typically have low stellar masses, $M_\star<10^7\mathrm{M}_\odot$, and star formation rates, SFR$<10^{-3}\mathrm{M}_\odot\mathrm{yr}^{-1}$.\\
(b) the {\it localized escape mode}  is characterized by compact ionized channels (with a Gini score of $0.93$) in central, highly star forming parts of galaxies with large scale heights ($H>3 \times 10^{21}$cm). It is mainly limited by dust opacity, resulting in a negligible LyC escape for galaxies with $Z>10^{-3}$ (given our assumption that dust content is linearly correlated with the metallicity). Galaxies leaking LyC through this mode have stellar masses $M_\star<10^8\mathrm{M}_\odot$ and star formation rates in the range $10^{-5} \mathrm{M}_\odot\mathrm{yr}^{-1} <\mathrm{SFR} < 1 \mathrm{M}_\odot\mathrm{yr}^{-1}$. 
    \item The average escape fraction decreases with increasing redshift for galaxies with $M_\star<10^7\mathrm{M}_\odot$. This behaviour is mainly driven by the \textit{ext}-mode, which represent the main LyC escape channel in low mass, relatively high metallicity galaxies, i.e. the typical properties of high redshift galaxies. This trend is however unlikely to continue to lower redshifts given the eventual quenching of low mass galaxies in the post-reionization era.
    \item We provide a handy fitting function that predicts the LyC escape fraction of a single galaxy within a factor of $\sim 2$, with a relative deviation of $3.4\%$ in the average escape fraction. Our model is statistically able to capture most features of LyC leakage as a function of galactic properties such as the bimodality, and the dependence of the escape fraction on redshift, SFR and stellar mass. This model is well-suited to describe the subgrid escape of galaxies for large-scale simulations.
\end{itemize}

While the model discussed in this paper has been proven effective at characterizing the escape of LyC radiation, in the future we plan to {\it (i)} test its validity on different simulations, {\it (ii)} extend it to take into account a more complex structure of the absorbing gas, and {\it (iii)} investigate the relative contribution of the two escape modes to reionization through cosmic time.

\section*{Data Availability}

Data directly related to this publication and its figures is available on request from the corresponding author. The IllustrisTNG simulations, including TNG50, are publicly available and accessible at \url{www.tng-project.org/data} \citep[see][]{nelson19a}.

\section*{Acknowledgements}
The authors extend their gratitude to Enrico Garaldi for insightful discussions that have contributed to the development of this work. Furthermore, we would like to express our appreciation to Dylan Nelson and Ruediger Pakmor for their valuable recommendations regarding the post-processing of TNG50 data.

\bibliographystyle{mnras}
\bibliography{refs}

\begin{thebibliography}{}
\makeatletter
\relax
\def\mn@urlcharsother{\let\do\@makeother \do\$\do\&\do\#\do\^\do\_\do\%\do\~}
\def\mn@doi{\begingroup\mn@urlcharsother \@ifnextchar [ {\mn@doi@} {\mn@doi@[]}}
\def\mn@doi@[#1]#2{\def\@tempa{#1}\ifx\@tempa\@empty \href {http://dx.doi.org/#2} {doi:#2}\else \href {http://dx.doi.org/#2} {#1}\fi \endgroup}
\def\mn@eprint#1#2{\mn@eprint@#1:#2::\@nil}
\def\mn@eprint@arXiv#1{\href {http://arxiv.org/abs/#1} {{\tt arXiv:#1}}}
\def\mn@eprint@dblp#1{\href {http://dblp.uni-trier.de/rec/bibtex/#1.xml} {dblp:#1}}
\def\mn@eprint@#1:#2:#3:#4\@nil{\def\@tempa {#1}\def\@tempb {#2}\def\@tempc {#3}\ifx \@tempc \@empty \let \@tempc \@tempb \let \@tempb \@tempa \fi \ifx \@tempb \@empty \def\@tempb {arXiv}\fi \@ifundefined {mn@eprint@\@tempb}{\@tempb:\@tempc}{\expandafter \expandafter \csname mn@eprint@\@tempb\endcsname \expandafter{\@tempc}}}

\bibitem[\protect\citeauthoryear{Alavi et~al.,}{Alavi et~al.}{2020}]{alavi20}
Alavi A.,  et~al., 2020, \mn@doi [The Astrophysical Journal] {10.3847/1538-4357/abbd43}, 904, 59

\bibitem[\protect\citeauthoryear{Asplund, Grevesse, Sauval  \& Scott}{Asplund et~al.}{2009}]{asplund09}
Asplund M.,  Grevesse N.,  Sauval A.~J.,   Scott P.,  2009, \mn@doi [Annual Review of Astronomy and Astrophysics] {10.1146/annurev.astro.46.060407.145222}, 47, 481

\bibitem[\protect\citeauthoryear{Boutsia et~al.,}{Boutsia et~al.}{2011}]{boutsia11}
Boutsia K.,  et~al., 2011, \mn@doi [The Astrophysical Journal] {10.1088/0004-637X/736/1/41}, 736, 41

\bibitem[\protect\citeauthoryear{Chabrier}{Chabrier}{2003}]{chabrier03}
Chabrier G.,  2003, Publications of the Astronomical Society of the Pacific, 115, 763

\bibitem[\protect\citeauthoryear{{Chisholm, J.}, {Orlitov\'a, I.}, {Schaerer, D.}, {Verhamme, A.}, {Worseck, G.}, {Izotov, Y. I.}, {Thuan, T. X.}  \& {Guseva, N. G.}}{{Chisholm, J.} et~al.}{2017}]{chisholm17}
{Chisholm, J.} {Orlitov\'a, I.} {Schaerer, D.} {Verhamme, A.} {Worseck, G.} {Izotov, Y. I.} {Thuan, T. X.}  {Guseva, N. G.} 2017, \mn@doi [A\&A] {10.1051/0004-6361/201730610}, 605, A67

\bibitem[\protect\citeauthoryear{Chisholm, Prochaska, Schaerer, Gazagnes  \& Henry}{Chisholm et~al.}{2020}]{chisholm20}
Chisholm J.,  Prochaska J.~X.,  Schaerer D.,  Gazagnes S.,   Henry A.,  2020, \mn@doi [Monthly Notices of the Royal Astronomical Society] {10.1093/mnras/staa2470}, 498, 2554

\bibitem[\protect\citeauthoryear{{Chisholm} et~al.,}{{Chisholm} et~al.}{2022}]{chisholm22}
{Chisholm} J.,  et~al., 2022, \mn@doi [Monthly Notices of the Royal Astronomical Society] {10.1093/mnras/stac2874}, \href {https://ui.adsabs.harvard.edu/abs/2022MNRAS.517.5104C} {517, 5104}

\bibitem[\protect\citeauthoryear{Ciardi, Ferrara, Marri  \& Raimondo}{Ciardi et~al.}{2001}]{crash}
Ciardi B.,  Ferrara A.,  Marri S.,   Raimondo G.,  2001, Monthly Notices of the Royal Astronomical Society, 324, 381

\bibitem[\protect\citeauthoryear{Davis, Efstathiou, Frenk  \& White}{Davis et~al.}{1985}]{davis85}
Davis M.,  Efstathiou G.,  Frenk C.~S.,   White S.~D.,  1985, The Astrophysical Journal, 292, 371

\bibitem[\protect\citeauthoryear{{Dayal} \& {Ferrara}}{{Dayal} \& {Ferrara}}{2018}]{dayal18}
{Dayal} P.,  {Ferrara} A.,  2018, \mn@doi [Physics Reports] {10.1016/j.physrep.2018.10.002}, \href {https://ui.adsabs.harvard.edu/abs/2018PhR...780....1D} {780, 1}

\bibitem[\protect\citeauthoryear{{Dayal} et~al.,}{{Dayal} et~al.}{2022}]{dayal22}
{Dayal} P.,  et~al., 2022, \mn@doi [Monthly Notices of the Royal Astronomical Society] {10.1093/mnras/stac537}, \href {https://ui.adsabs.harvard.edu/abs/2022MNRAS.512..989D} {512, 989}

\bibitem[\protect\citeauthoryear{Dijkstra, Wyithe, Haiman, Mesinger  \& Pentericci}{Dijkstra et~al.}{2014}]{dijkstra14}
Dijkstra M.,  Wyithe S.,  Haiman Z.,  Mesinger A.,   Pentericci L.,  2014, \mn@doi [Monthly Notices of the Royal Astronomical Society] {10.1093/mnras/stu531}, 440, 3309

\bibitem[\protect\citeauthoryear{{Faucher-Gigu{\`e}re}, {Lidz}, {Zaldarriaga}  \& {Hernquist}}{{Faucher-Gigu{\`e}re} et~al.}{2009}]{fg09}
{Faucher-Gigu{\`e}re} C.-A.,  {Lidz} A.,  {Zaldarriaga} M.,   {Hernquist} L.,  2009, \mn@doi [\apj] {10.1088/0004-637X/703/2/1416}, \href {http://adsabs.harvard.edu/abs/2009ApJ...703.1416F} {703, 1416}

\bibitem[\protect\citeauthoryear{Ferrara, Vallini, Pallottini, Gallerani, Carniani, Kohandel, Decataldo  \& Behrens}{Ferrara et~al.}{2019}]{ferrara19}
Ferrara A.,  Vallini L.,  Pallottini A.,  Gallerani S.,  Carniani S.,  Kohandel M.,  Decataldo D.,   Behrens C.,  2019, \mn@doi [Monthly Notices of the Royal Astronomical Society] {10.1093/mnras/stz2031}, 489, 1

\bibitem[\protect\citeauthoryear{{Grazian} et~al.,}{{Grazian} et~al.}{2016}]{grazian16}
{Grazian} A.,  et~al., 2016, \mn@doi [Astronomy and Astrophysics] {10.1051/0004-6361/201526396}, \href {https://ui.adsabs.harvard.edu/abs/2016A&A...585A..48G} {585, A48}

\bibitem[\protect\citeauthoryear{Graziani, Maselli  \& Ciardi}{Graziani et~al.}{2013}]{crash3}
Graziani L.,  Maselli A.,   Ciardi B.,  2013, Monthly Notices of the Royal Astronomical Society, 431, 722

\bibitem[\protect\citeauthoryear{{Heckman}, {Sembach}, {Meurer}, {Leitherer}, {Calzetti}  \& {Martin}}{{Heckman} et~al.}{2001}]{heckman01}
{Heckman} T.~M.,  {Sembach} K.~R.,  {Meurer} G.~R.,  {Leitherer} C.,  {Calzetti} D.,   {Martin} C.~L.,  2001, \mn@doi [Astrophysical Journal] {10.1086/322475}, \href {https://ui.adsabs.harvard.edu/abs/2001ApJ...558...56H} {558, 56}

\bibitem[\protect\citeauthoryear{Inoue, Iwata  \& Deharveng}{Inoue et~al.}{2006}]{inoue06}
Inoue A.~K.,  Iwata I.,   Deharveng J.-M.,  2006, \mn@doi [Monthly Notices of the Royal Astronomical Society: Letters] {10.1111/j.1745-3933.2006.00195.x}, 371, L1

\bibitem[\protect\citeauthoryear{Jones, Ellis, Schenker  \& Stark}{Jones et~al.}{2013}]{jones13}
Jones T.~A.,  Ellis R.~S.,  Schenker M.~A.,   Stark D.~P.,  2013, \mn@doi [The Astrophysical Journal] {10.1088/0004-637X/779/1/52}, 779, 52

\bibitem[\protect\citeauthoryear{{Katz} et~al.,}{{Katz} et~al.}{2023}]{katz22b}
{Katz} H.,  et~al., 2023, \mn@doi [\mnras] {10.1093/mnras/stac3019}, \href {https://ui.adsabs.harvard.edu/abs/2023MNRAS.518..270K} {518, 270}

\bibitem[\protect\citeauthoryear{Kimm \& Cen}{Kimm \& Cen}{2014}]{kimm14}
Kimm T.,  Cen R.,  2014, \mn@doi [Astrophysical Journal] {10.1088/0004-637X/788/2/121}, 788

\bibitem[\protect\citeauthoryear{{Kostyuk}, {Nelson}, {Ciardi}, {Glatzle}  \& {Pillepich}}{{Kostyuk} et~al.}{2023}]{kostyuk23}
{Kostyuk} I.,  {Nelson} D.,  {Ciardi} B.,  {Glatzle} M.,   {Pillepich} A.,  2023, \mn@doi [Monthly Notices of the Royal Astronomical Society] {10.1093/mnras/stad677}, \href {https://ui.adsabs.harvard.edu/abs/2023MNRAS.521.3077K} {521, 3077}

\bibitem[\protect\citeauthoryear{{Leitet}, {Bergvall}, {Piskunov}  \& {Andersson}}{{Leitet} et~al.}{2011}]{leitet11}
{Leitet} E.,  {Bergvall} N.,  {Piskunov} N.,   {Andersson} B.~G.,  2011, \mn@doi [Astronomy and Astrophysics] {10.1051/0004-6361/201015654}, \href {https://ui.adsabs.harvard.edu/abs/2011A&A...532A.107L} {532, A107}

\bibitem[\protect\citeauthoryear{Leitherer, Schaerer, Goldader, Gonza, Delgado, Kune, Devost  \& Heckman}{Leitherer et~al.}{1999}]{leitherer99}
Leitherer C.,  Schaerer D.,  Goldader J.~D.,  Gonza R.~M.,  Delgado L. E.~Z.,  Kune D. F. O.~O.,  Devost D.,   Heckman T.~M.,  1999, ApJS, pp 3--40

\bibitem[\protect\citeauthoryear{Lewis et~al.,}{Lewis et~al.}{2020}]{lewis20}
Lewis J.~S.,  et~al., 2020, \mn@doi [Monthly Notices of the Royal Astronomical Society] {10.1093/mnras/staa1748}, 496, 4342

\bibitem[\protect\citeauthoryear{Ma, Hopkins, Kasen, Quataert, Faucher-Gigu{\`{e}}re, Kere{\v{s}}, Murray  \& Strom}{Ma et~al.}{2016}]{ma16}
Ma X.,  Hopkins P.~F.,  Kasen D.,  Quataert E.,  Faucher-Gigu{\`{e}}re C.-A.,  Kere{\v{s}} D.,  Murray N.,   Strom A.,  2016, \mn@doi [Monthly Notices of the Royal Astronomical Society] {10.1093/mnras/stw941}, 459, 3614

\bibitem[\protect\citeauthoryear{Ma, Quataert, Wetzel, Hopkins, Faucher-Gigu{\`{e}}re  \& Kere{\v{s}}}{Ma et~al.}{2020}]{ma20}
Ma X.,  Quataert E.,  Wetzel A.,  Hopkins P.~F.,  Faucher-Gigu{\`{e}}re C.~A.,   Kere{\v{s}} D.,  2020, \mn@doi [Monthly Notices of the Royal Astronomical Society] {10.1093/mnras/staa2404}, 498, 2001

\bibitem[\protect\citeauthoryear{{Marinacci} et~al.,}{{Marinacci} et~al.}{2018}]{marinacci18}
{Marinacci} F.,  et~al., 2018, \mn@doi [\mnras] {10.1093/mnras/sty2206}, \href {https://ui.adsabs.harvard.edu/abs/2018MNRAS.480.5113M} {480, 5113}

\bibitem[\protect\citeauthoryear{Maselli, Ferrara  \& Ciardi}{Maselli et~al.}{2003}]{crash_upgrade}
Maselli A.,  Ferrara A.,   Ciardi B.,  2003, Monthly Notices of the Royal Astronomical Society, 345, 379

\bibitem[\protect\citeauthoryear{Maselli, Ciardi  \& Kanekar}{Maselli et~al.}{2009}]{crash2}
Maselli A.,  Ciardi B.,   Kanekar A.,  2009, Monthly Notices of the Royal Astronomical Society, 393, 171

\bibitem[\protect\citeauthoryear{Matthee, Sobral, Best, Khostovan, Oteo, Bouwens  \& Röttgering}{Matthee et~al.}{2016}]{mathee16}
Matthee J.,  Sobral D.,  Best P.,  Khostovan A.~A.,  Oteo I.,  Bouwens R.,   Röttgering H.,  2016, \mn@doi [Monthly Notices of the Royal Astronomical Society] {10.1093/mnras/stw2973}, 465, 3637

\bibitem[\protect\citeauthoryear{Mostardi, Shapley, Nestor, Steidel, Reddy  \& Trainor}{Mostardi et~al.}{2013}]{mostardi13}
Mostardi R.~E.,  Shapley A.~E.,  Nestor D.~B.,  Steidel C.~C.,  Reddy N.~A.,   Trainor R.~F.,  2013, \mn@doi [The Astrophysical Journal] {10.1088/0004-637X/779/1/65}, 779, 65

\bibitem[\protect\citeauthoryear{{Naiman} et~al.,}{{Naiman} et~al.}{2018}]{naiman18}
{Naiman} J.~P.,  et~al., 2018, \mn@doi [\mnras] {10.1093/mnras/sty618}, \href {https://ui.adsabs.harvard.edu/abs/2018MNRAS.477.1206N} {477, 1206}

\bibitem[\protect\citeauthoryear{{Nakajima} \& {Ouchi}}{{Nakajima} \& {Ouchi}}{2014}]{nakajima14}
{Nakajima} K.,  {Ouchi} M.,  2014, \mn@doi [Monthly Notices of the Royal Astronomical Society] {10.1093/mnras/stu902}, \href {https://ui.adsabs.harvard.edu/abs/2014MNRAS.442..900N} {442, 900}

\bibitem[\protect\citeauthoryear{{Nelson} et~al.,}{{Nelson} et~al.}{2018}]{nelson18a}
{Nelson} D.,  et~al., 2018, \mn@doi [\mnras] {10.1093/mnras/stx3040}, \href {https://ui.adsabs.harvard.edu/abs/2018MNRAS.475..624N} {475, 624}

\bibitem[\protect\citeauthoryear{{Nelson} et~al.,}{{Nelson} et~al.}{2019}]{nelson19a}
{Nelson} D.,  et~al., 2019, \mn@doi [Computational Astrophysics and Cosmology] {10.1186/s40668-019-0028-x}, \href {https://ui.adsabs.harvard.edu/abs/2019ComAC...6....2N} {6, 2}

\bibitem[\protect\citeauthoryear{{Nestor}, {Shapley}, {Steidel}  \& {Siana}}{{Nestor} et~al.}{2011}]{nestor11}
{Nestor} D.~B.,  {Shapley} A.~E.,  {Steidel} C.~C.,   {Siana} B.,  2011, \mn@doi [The Astrophysical Journal] {10.1088/0004-637X/736/1/18}, \href {https://ui.adsabs.harvard.edu/abs/2011ApJ...736...18N} {736, 18}

\bibitem[\protect\citeauthoryear{Ocvirk et~al.,}{Ocvirk et~al.}{2016}]{ocvirk16}
Ocvirk P.,  et~al., 2016, \mn@doi [Monthly Notices of the Royal Astronomical Society] {10.1093/mnras/stw2036}, 463, 1462

\bibitem[\protect\citeauthoryear{Ocvirk, Lewis, Gillet, Chardin, Aubert, Deparis  \& Th{\'{e}}lie}{Ocvirk et~al.}{2021}]{ocvirk21}
Ocvirk P.,  Lewis J.~S.,  Gillet N.,  Chardin J.,  Aubert D.,  Deparis N.,   Th{\'{e}}lie E.,  2021, \mn@doi [Monthly Notices of the Royal Astronomical Society] {10.1093/mnras/stab2502}, 507, 6108

\bibitem[\protect\citeauthoryear{Paardekooper, Khochfar  \& Vecchia}{Paardekooper et~al.}{2015}]{paardekooper15}
Paardekooper J.~P.,  Khochfar S.,   Vecchia C.~D.,  2015, \mn@doi [Monthly Notices of the Royal Astronomical Society] {10.1093/mnras/stv1114}, 451, 2544

\bibitem[\protect\citeauthoryear{Pahl, Shapley, Steidel, Chen  \& Reddy}{Pahl et~al.}{2021}]{pahl21}
Pahl A.~J.,  Shapley A.,  Steidel C.~C.,  Chen Y.,   Reddy N.~A.,  2021, \mn@doi [Monthly Notices of the Royal Astronomical Society] {10.1093/mnras/stab1374}, 505, 2447

\bibitem[\protect\citeauthoryear{{Pillepich} et~al.,}{{Pillepich} et~al.}{2018a}]{pillepich18a}
{Pillepich} A.,  et~al., 2018a, \mn@doi [\mnras] {10.1093/mnras/stx2656}, \href {https://ui.adsabs.harvard.edu/abs/2018MNRAS.473.4077P} {473, 4077}

\bibitem[\protect\citeauthoryear{{Pillepich} et~al.,}{{Pillepich} et~al.}{2018b}]{pillepich18b}
{Pillepich} A.,  et~al., 2018b, \mn@doi [\mnras] {10.1093/mnras/stx3112}, \href {https://ui.adsabs.harvard.edu/abs/2018MNRAS.475..648P} {475, 648}

\bibitem[\protect\citeauthoryear{{Pillepich} et~al.,}{{Pillepich} et~al.}{2019}]{pillepich19}
{Pillepich} A.,  et~al., 2019, \mn@doi [\mnras] {10.1093/mnras/stz2338}, \href {https://ui.adsabs.harvard.edu/abs/2019MNRAS.490.3196P} {490, 3196}

\bibitem[\protect\citeauthoryear{{Planck Collaboration}}{{Planck Collaboration}}{2016}]{planck16}
{Planck Collaboration} 2016, \mn@doi [\aap] {10.1051/0004-6361/201527101}, \href {https://ui.adsabs.harvard.edu/abs/2016A&A...594A...1P} {594, A1}

\bibitem[\protect\citeauthoryear{Rosdahl et~al.,}{Rosdahl et~al.}{2018}]{rosdahl18}
Rosdahl J.,  et~al., 2018, \mn@doi [Monthly Notices of the Royal Astronomical Society] {10.1093/mnras/sty1655}, 479, 994

\bibitem[\protect\citeauthoryear{Rosdahl et~al.,}{Rosdahl et~al.}{2022}]{rosdahl22}
Rosdahl J.,  et~al., 2022, {LyC escape from SPHINX galaxies in the Epoch of Reionization} (\mn@eprint {arXiv} {2207.03232}), \url {http://arxiv.org/abs/2207.03232}

\bibitem[\protect\citeauthoryear{Springel}{Springel}{2010}]{springel10}
Springel V.,  2010, Monthly Notices of the Royal Astronomical Society, 401, 791

\bibitem[\protect\citeauthoryear{Springel \& Hernquist}{Springel \& Hernquist}{2003}]{springel03}
Springel V.,  Hernquist L.,  2003, Monthly Notices of the Royal Astronomical Society, 339, 289

\bibitem[\protect\citeauthoryear{Springel, White, Tormen  \& Kauffmann}{Springel et~al.}{2001}]{springel01}
Springel V.,  White S.~D.,  Tormen G.,   Kauffmann G.,  2001, Monthly Notices of the Royal Astronomical Society, 328, 726

\bibitem[\protect\citeauthoryear{Springel et~al.,}{Springel et~al.}{2005}]{springel05}
Springel V.,  et~al., 2005, Nature, 435, 629

\bibitem[\protect\citeauthoryear{{Springel} et~al.,}{{Springel} et~al.}{2018}]{springel18}
{Springel} V.,  et~al., 2018, \mn@doi [\mnras] {10.1093/mnras/stx3304}, \href {https://ui.adsabs.harvard.edu/abs/2018MNRAS.475..676S} {475, 676}

\bibitem[\protect\citeauthoryear{Tang, Stark, Chevallard, Charlot, Endsley  \& Congiu}{Tang et~al.}{2021}]{tang21}
Tang M.,  Stark D.,  Chevallard J.,  Charlot S.,  Endsley R.,   Congiu E.,  2021, \mn@doi [Monthly Notices of the Royal Astronomical Society] {10.1093/mnras/stab705}, 503

\bibitem[\protect\citeauthoryear{Vanzella et~al.,}{Vanzella et~al.}{2010}]{vanzella10}
Vanzella E.,  et~al., 2010, \mn@doi [The Astrophysical Journal] {10.1088/0004-637X/725/1/1011}, 725, 1011

\bibitem[\protect\citeauthoryear{{Verhamme}, {Orlitov{\'a}}, {Schaerer}  \& {Hayes}}{{Verhamme} et~al.}{2015}]{verhamme15}
{Verhamme} A.,  {Orlitov{\'a}} I.,  {Schaerer} D.,   {Hayes} M.,  2015, \mn@doi [\aap] {10.1051/0004-6361/201423978}, \href {https://ui.adsabs.harvard.edu/abs/2015A&A...578A...7V} {578, A7}

\bibitem[\protect\citeauthoryear{{Weinberger} et~al.,}{{Weinberger} et~al.}{2017}]{weinberger17}
{Weinberger} R.,  et~al., 2017, \mn@doi [\mnras] {10.1093/mnras/stw2944}, \href {https://ui.adsabs.harvard.edu/abs/2017MNRAS.465.3291W} {465, 3291}

\bibitem[\protect\citeauthoryear{Weingartner \& Draine}{Weingartner \& Draine}{2001}]{weingartner01}
Weingartner J.~C.,  Draine B.~T.,  2001, \mn@doi [The Astrophysical Journal] {10.1086/318651}, 548, 296

\bibitem[\protect\citeauthoryear{Xu, Wise, Norman, Ahn  \& O'Shea}{Xu et~al.}{2016}]{xu16}
Xu H.,  Wise J.~H.,  Norman M.~L.,  Ahn K.,   O'Shea B.~W.,  2016, \mn@doi [The Astrophysical Journal] {10.3847/1538-4357/833/1/84}, 833, 84

\bibitem[\protect\citeauthoryear{{Yeh} et~al.,}{{Yeh} et~al.}{2023}]{yeh22}
{Yeh} J. Y.~C.,  et~al., 2023, \mn@doi [\mnras] {10.1093/mnras/stad210}, \href {https://ui.adsabs.harvard.edu/abs/2023MNRAS.520.2757Y} {520, 2757}

\makeatother
\end{thebibliography}

\appendix
\section{Convergence with grid size}  \label{app:convergence_test}
In order to accurately capture the escape of LyC radiation, we need to ensure that the galaxies are sufficiently resolved. In our approach, we distribute the gas particles of the TNG50 galaxy, as well as their physical properties, on a regular Carthesian grid before projecting all galactic properties as surface densities onto the galactic plane. To ensure the escape fraction we obtain is sufficiently resolved, we applied this method to a sample of galaxies varying the grid sizes in the range of $2^3$ to $502^3$ in steps of 20. We used a sample of 80 galaxies. To ensure that the sample represents the range of galaxies in the TNG50 simulation, the halos were sampled from logarithmically spaced stellar mass bins, ranging from the least to the most massive galaxies in the simulation.

\begin{figure}
    \centering
    \includegraphics[width=0.49\textwidth]{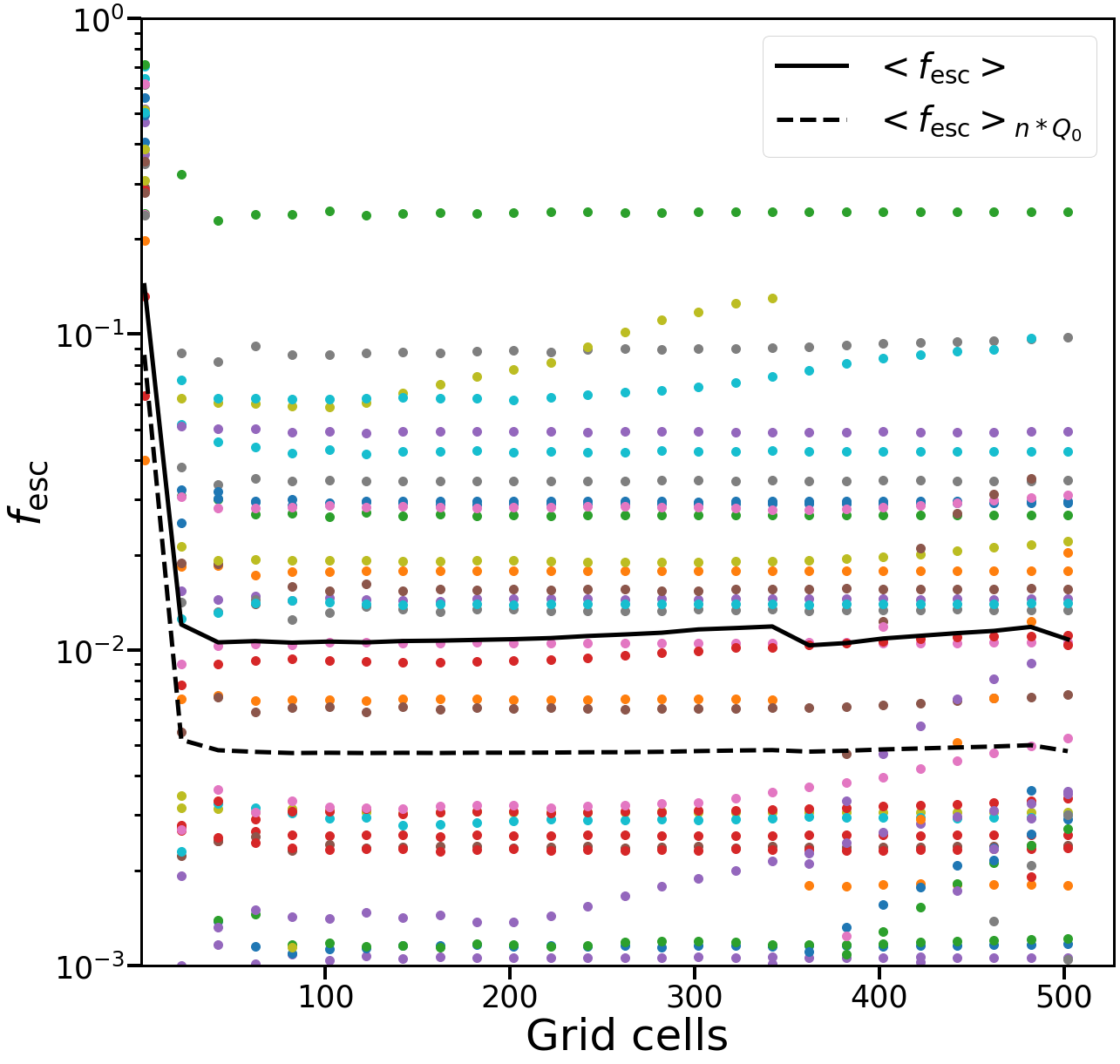}
    \caption{Escape fraction of a sample of galaxies as a function of the number of grid cells per dimension. We additionally depict the average escape fraction (black continuous line) and the average escape fraction weighted by the emissivity (black dashed line).}
    \label{fig:gridpoint_variation}
\end{figure}

The result of the convergence study can be seen in fig.~\ref{fig:gridpoint_variation}. We observe that $f_\mathrm{esc}$ rapidly converges with grid size, i.e. there is no significant change in $f_\mathrm{esc}$ for grid sizes $>60^3$. The increase in $f_\mathrm{esc}$ for small grid sizes is explained by the smoothing out of regions containing dense gas, thus decreasing its optical depth. Finally, at grid sizes $>150^3$ in some galaxies the escape fraction increases, while for a few others it becomes undefined. This is likely caused by numerical problems as in some grid cells the smoothing kernel decreases to small, numerically unstable values. Based on these results we have chosen $100^3$ grid cells to be a reliable number to ensure that the LyC escape fraction is fully resolved within the resolution limits of the TNG50 simulation itself.

\section{Extent of the star forming regions} \label{app:sf-scale}
As discussed previously, our model assumes that star formation, and hence LyC production, takes place in a thin slab. This assumption is based on the fact that most star formation is confined to a region significantly narrower than the one occupied by the gas which inhibits radiation escape. Consequently, a substantial portion of LyC photons is required to traverse the majority of the intervening gas.

\begin{figure}
    \centering
    \includegraphics[width=0.49\textwidth]{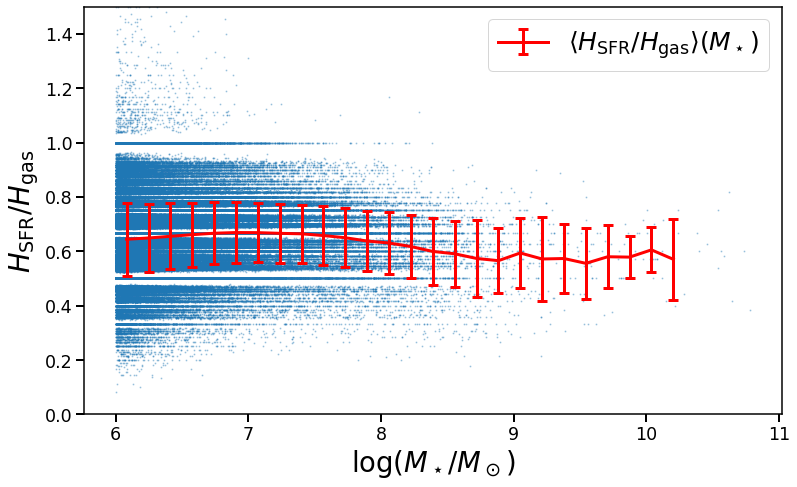}
    \caption{Ratio between the scale height of the star forming region and that of the gas as a function of galactic stellar mass. The red line represents the average value of this ratio, and the errorbars its one sigma scatter. The linear pattern in which the points appear is a consequence of measuring the scale height in discrete grid cell numbers.}
    \label{fig:sfr_vs_gas}
\end{figure}

While this intuitively holds for large, disc-shaped galaxies, its validity may be questioned when applied to smaller galaxies that exhibit a more spherical morphology. In order to verify this assumption, we examine the ratio between the scale height of the star forming region ($H_\mathrm{SFR}$) and that of the gas ($H_\mathrm{gas}$), as illustrated in fig.~\ref{fig:sfr_vs_gas}.

Our analysis reveals an average ratio of approximately $0.6$, implying that in most galaxies, on average, the star forming region is considerably closer to the center than the majority of the gas. Although this value suggests that our model might overlook some potential LyC escape originating from star forming regions positioned above the galactic plane, given the qualitative and statistical nature of our study, the model provides a reasonable estimate of the likelihood of LyC radiation escaping.

Notably, we see a lack of substantial evolution in the ratio with increasing stellar mass. Indeed, galaxies with higher stellar masses tend collapse into a more disc-like structure, i.e. a lower scale height of the star forming region. However, the gas enveloping the galaxy is also more collapsed, resulting in a ratio which is not too dissimilar from that of smaller galaxies.
%What becomes evident is that the collapse of both the star-forming region and the surrounding gas evolves at a comparable rate with the stellar mass of the galaxy. 

Although a larger number of galaxies exhibit higher values of $H_\mathrm{SFR}/H_\mathrm{gas}$ for smaller $M_\star$, this pattern arises solely from the larger sample size. This is evident as the width of the 1-sigma scatter of $H_\mathrm{SFR}/H_\mathrm{gas}$ values remains relatively constant across the entire stellar mass range.

Summarizing, despite the fact that smaller galaxies have a much more diffuse structure, the thin slate assumption should still hold for the galaxy population as a whole (although this is not the case for single galaxies) given that the LyC producing region is significantly thinner than the distribution of the gas inhibiting its escape.

\section{Comparison with numerical post-processing} \label{app:comparison_num}
In \cite{kostyuk23} we investigated the escape fraction of $\sim 10^4$ galaxies from TNG50 using the Monte-Carlo RT code CRASH \citep{crash, crash2, crash_upgrade, crash3}. While there is an overlap in the galaxies that we have examined the methodology significantly differs. 

\begin{figure}
    \includegraphics[width=0.24\textwidth]{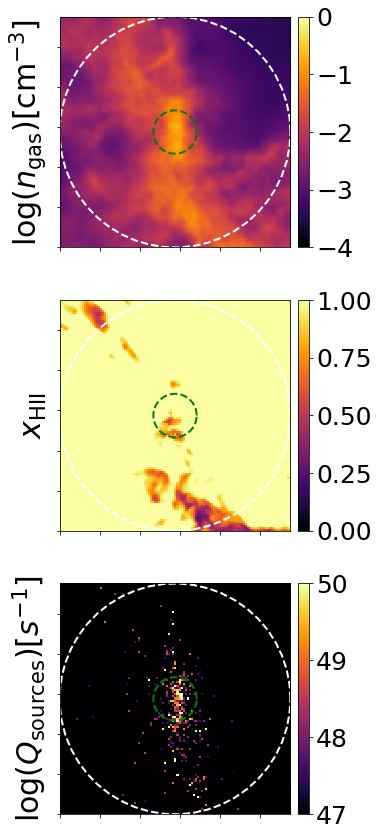}
    \includegraphics[width=0.24\textwidth]{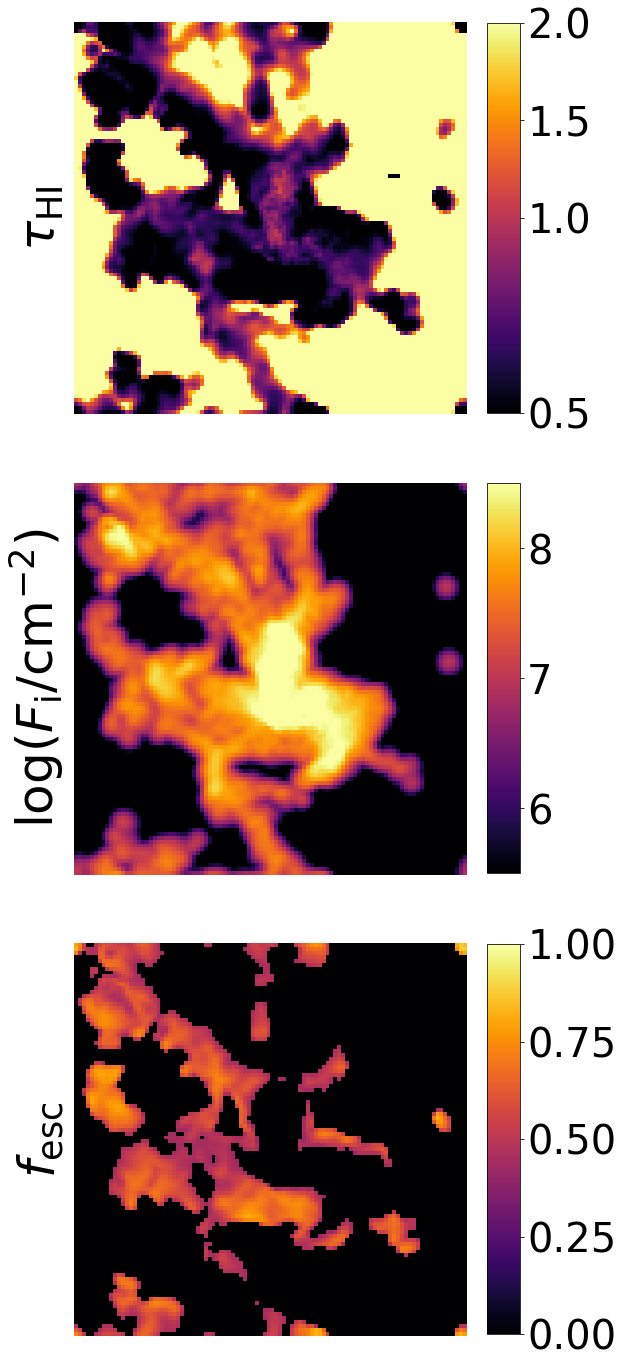}
    
    \caption{Comparison between a galaxy processed with CRASH-RT \textit{(left)} and with the semianalytic model \textit{(right)}. On the left we depict from top to bottom the average gas density ($n_\mathrm{gas}$), the average ionization fraction ($x_\mathrm{HII}$) and the distribution and ionizing emissivity of sources ($Q_\mathrm{sources}$). On the right we depict the hydrogen optical depth ($\tau_\mathrm{HI}$), ionizing emissivity $F_\mathrm{ion}$ and the location based escape fraction ($f_\mathrm{esc}$).
    }
    \label{fig:example_galaxy}
\end{figure}

In fig.~\ref{fig:example_galaxy} we depict projections of one example galaxy which was examined in both \cite{kostyuk23} and this work. In the former study the escape fraction was determined to be $f_\mathrm{esc,RT}=63.5\%$ while in this study it yielded $f_\mathrm{esc}=12.7\%$. We note that the orientation of the galaxy differs as the gas in the RT case was gridded with respect to its TNG50 coordinates as photon emission takes place isotropically in the simulation and is therefore orientation invariant, while, as described in section \ref{sec:semi_method}, in the semianalytic case the galaxy was first rotated into the galactic plane.

Given that we use different properties to calculate the escape fraction the characteristics that we depict in the plots are naturally also different. The relevant quantities to determine the escape fraction in the RT study were the gas density, the ionization fraction of that gas and finally the distribution of the ionizing sources. We see that while most of the gas is ionized, the escape fraction is reduced by neutral gas patches in the lower-half of central region as well as in the lower center of the galaxy. However, even in the central region most of the gas is ionized allowing for ionizing radiation to escape from the area of its highest production.

In the semianalytic study we see that the ionizing flux derived from the star-formation rate does approximately trace the distribution of ionizing sources on the left. However, as one can see by the optical depth and escape fraction distribution, ionizing photons are fully absorbed in the central region of the galaxy thus resulting in a much lower escape fraction as in the RT approach. It is likely that the discrepancy can be attributed to the smoothing out of the production area of ionizing photons and the resulting increase of the gas to photon ratio.

\begin{figure}
    \centering
    \includegraphics[width=0.49\textwidth]{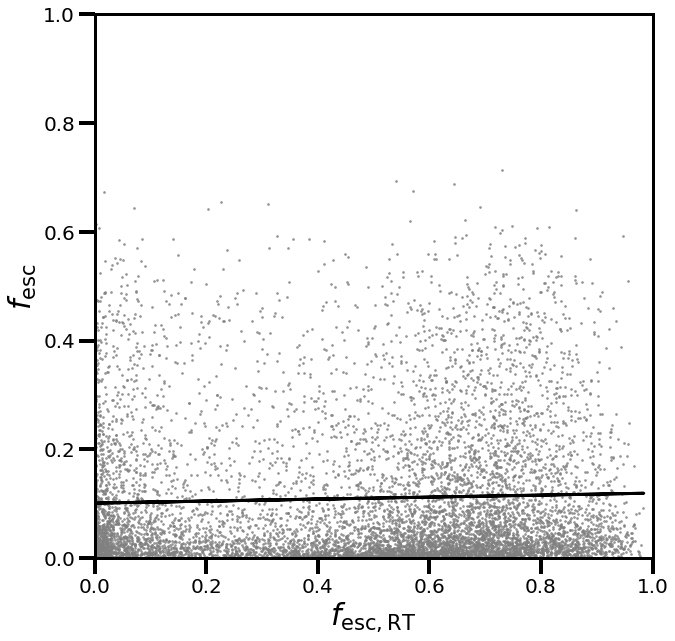}
    \caption{Escape fractions obtained using the analytic method discussed here, $f_\mathrm{esc}$, as a function of those obtained by post-processing the galaxies with the RT code CRASH, $f_\mathrm{esc, RT}$. The line represents a linear fit to the data.
    }
    \label{fig:num_comparison_fesc}
\end{figure}

Next we inspect the escape fractions of all galaxies processed with our two methods. In fig.~\ref{fig:num_comparison_fesc} we plot the escape fractions obtained here, $f_{\rm esc}$, as a function of those evaluated in \cite{kostyuk23}, $f_{\rm esc,RT}$. Surprisingly, only for $f_\mathrm{esc, RT}>0.6$ a slight correlation between the values obtained in the two studies can be seen. 
We believe that the reason for this behaviour is the different methodology used for obtaining the ionizing emissivity of a galaxy. Indeed, in the numerical approach this was determined for individual stellar particles representing a whole stellar population, since CRASH evaluates the propagation of photons emitted by the discrete sources. Moreover, the information encapsulated in the stellar particles, such as age, metallicity, and mass, allowed us to compute the full ionizing spectrum, enabling an investigation of LyC escape in individual frequency bins.
On the other hand, the SFR is a better tracer of the location of new stars, which are expected to emit most ionizing radiation. Additionally, a stellar particle-based approach in our semi-analytic model would lack numerical stability under grid refinement. Indeed, because of the fixed emissivity of star particles, a smaller grid-size would always result in a larger flux in cells hosting such particles.

\begin{figure}
    \centering
    \includegraphics[width=0.49\textwidth]{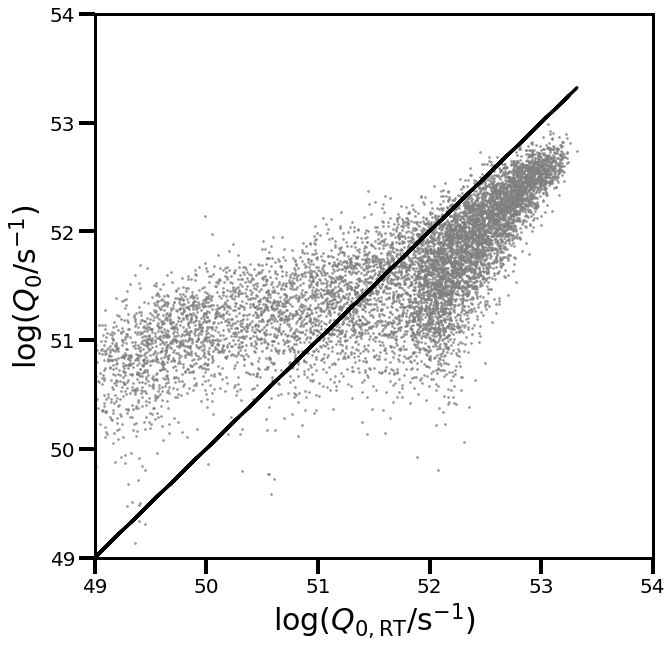}
    \caption{Ionizing emissivity  obtained from the SFR of galaxies, $Q_{0}$, as a function of that calculated using the spectra of stellar particles, $Q_{0,\mathrm{RT}}$. The line represents a 1:1 correspondence.}
    \label{fig:num_comparison}
\end{figure}
The difference introduced by the two approaches can be appreciated in fig~\ref{fig:num_comparison}, where the emissivities for a given galaxy obtained with the stellar particle, $Q_{\rm 0, RT}$, and with the SFR, $Q_{\rm 0}$, are plotted. It is evident that the scatter is very large, in particular for $Q_{\rm 0, RT}<10^{52}\mathrm{s}^{-1}$, where more than one order of magnitude difference is observed. The reason for this is that smaller galaxies contain very few stellar particles, such that random fluctuations in the spawning of one young stellar particle induce a large difference in the ionizing emissivity. The effect becomes less visible in larger galaxies, explaining the stronger correlation seen for $Q_{\rm 0, RT}>10^{52}\mathrm{s}^{-1}$.

\begin{figure}
    \centering
    \includegraphics[width=0.49\textwidth]{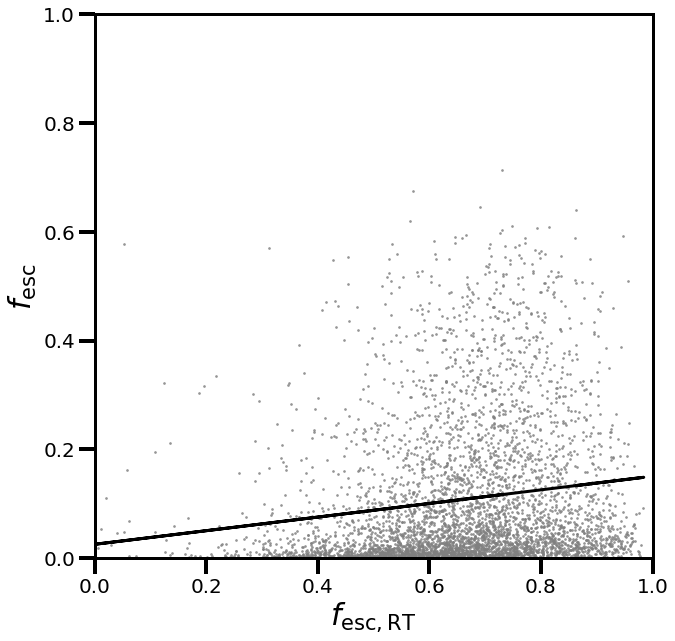}
    \caption{Same as fig.~\ref{fig:num_comparison_fesc} but  only for galaxies with $Q_{0,\mathrm{RT}}>10^{52}$~s$^{-1}$. The line represents a linear fit to the data.}
    \label{fig:num_comparison2}
\end{figure}
This is reflected in fig~\ref{fig:num_comparison2}, where we consider only the escape fraction of galaxies with $Q_{\rm 0, RT}>10^{52}\mathrm{s}^{-1}$, for which the correlation between the two escape fractions is much stronger. However, the scatter is still significant due to the geometry of the problem. 
Indeed, here the full star forming regions are considered as source of radiation, meaning that LyC emission is less concentrated and thus less sensitive to the fluctuation in the densities of the surrounding gas. This is not the case for stellar particles, when the escape fraction can change significantly because of random fluctuations in the relative position of a stellar particle and the surrounding gas. 
Finally, differently from the analytic approach, the numerical study models the radiation transfer of ionizing radiation.

\begin{figure}
    \centering
    \includegraphics[width=0.49\textwidth]{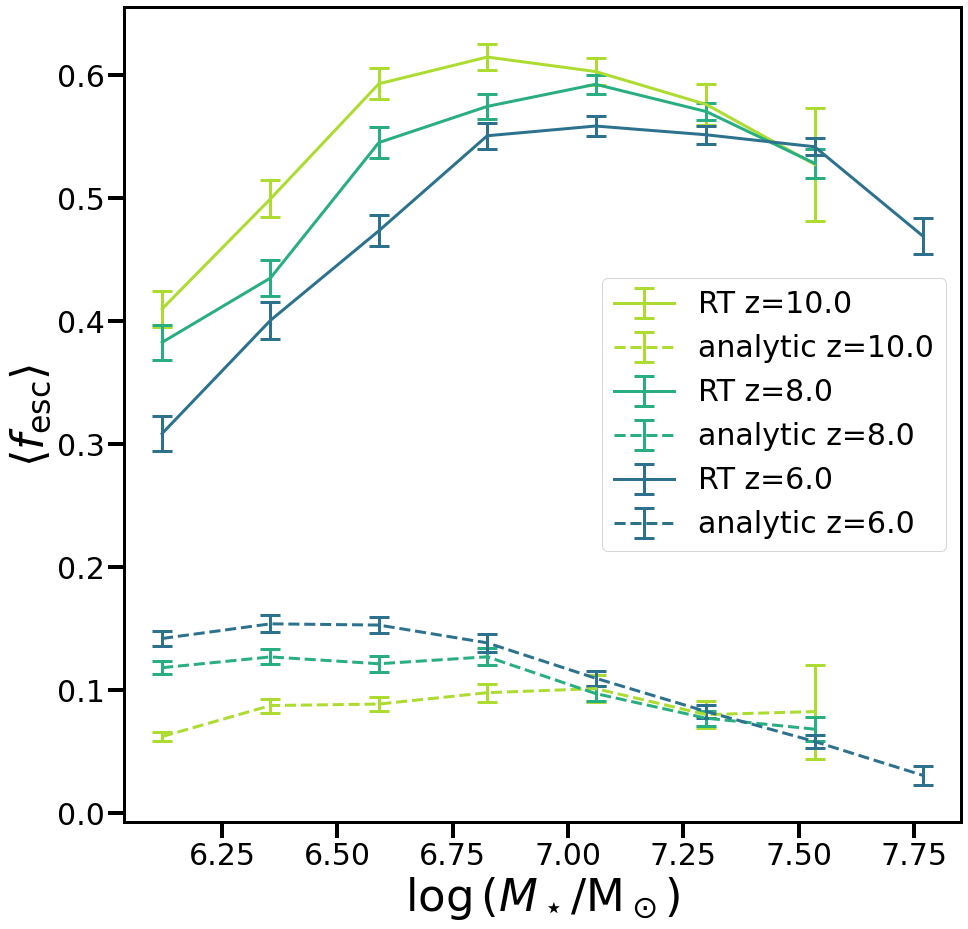}
    \caption{Comparison of $f_\mathrm{esc}$ dependence on stellar mass for the RT and semianalytic approaches.}
    \label{fig:fesc_Mstar_comp}
\end{figure}

Finally, in fig.~\ref{fig:fesc_Mstar_comp} we show the evolution of $f_\mathrm{esc}$ with stellar mass for the two approaches. Besides the main quantitative discrepancy in the values of the escape fraction we see two main differences in these results. Namely, the peak escape fraction is located at significantly higher stellar mass values in case of the RT approach. Most importantly however is the different evolution with redshift. As discussed earlier this discrepancy is likely to be caused by the \text{ext}-mode which emerges in the semianalytic model but is not present in the RT model due to the Poissonian sampling of the emitters of ionizing radiation.

\end{document}